\documentclass[twocolumn,prl,amsmath,amssymb,superscriptaddress]{revtex4-1}
\usepackage{graphicx,bm,dsfont,amsmath,amssymb}
\usepackage{soul}
\usepackage{color}
\usepackage{amsthm}
\usepackage{mathrsfs}
\usepackage{subfigure}
\usepackage[colorlinks,linkcolor=blue,citecolor=blue,urlcolor=blue]{hyperref}
\newcommand{\blue}[1]{\textcolor{blue}{#1}}

\begin{document}
\title{Chiral Chaos Enhanced Sensing}
\author{Yun-Qiu Ge}
\thanks{\blue{These authors contribute equally to this work}}
\affiliation{School of integrated circuits, Tsinghua University, Beijing 100084, China}
\affiliation{Frontier Science Center for Quantum Information, Beijing, China}
\author{Zhe Wang}
\thanks{\blue{These authors contribute equally to this work}}
\affiliation{Department of Automation, Tsinghua University, Beijing 100084, China}
\author{Qian-Chuan Zhao}
\affiliation{Department of Automation, Tsinghua University, Beijing 100084, China}
\author{Jing Zhang}\email{zhangjing2022@xjtu.edu.cn}
\affiliation{School of Automation Science and Engineering, Xi'an Jiaotong University, Xi'an 710049, China}
\affiliation{MOE Key Lab for Intelligent Networks and Network Security, Xi'an Jiaotong University, Xi'an 710049, China}
\author{Yu-xi Liu}\email{yuxiliu@mail.tsinghua.edu.cn}
\affiliation{School of integrated circuits, Tsinghua University, Beijing 100084, China}
\affiliation{Frontier Science Center for Quantum Information, Beijing, China}

\begin{abstract}	
Chirality refers to the property that an object and its mirror image cannot overlap each other by spatial rotation and translation, and can be found in various research fields. We here propose chiral chaos and construct a chiral chaotic device via coupled whispering gallery mode resonators, where routes to chaos exhibit pronounced chirality for two opposite pumping directions. The mechanism responsible for this phenomenon is that time-reversal symmetry of the traveling-wave light fields is broken by the Rayleigh scatterers inserted in resonators. Combining with the Lyapunov exponents, we propose metrics to measure the symmetry and chirality between different chaotic dynamics. We find that such a chiral chaotic device can be applied to achieve sensing with high sensitivity, wide detectable range, and strong robustness to the phase and orientation randomness of weak signals. Our work presents a promising candidate for on-chip sensing and may have applications in quantum networks and chaotic communications.
\end{abstract}
\maketitle

\textit{Introduction}.---The concept of chirality has permeated into various research fields, e.g., soft matter physics~\cite{Jmaterchem2006}, chemistry~\cite{IntJBiomedSci}, biological science~\cite{Philo2016}, particle physics~\cite{RMP922020}, and optics~\cite{Nature5412017,RMP872015}. Recently, considerable efforts have been devoted to the research of chirality in light-matter inter-actions~\cite{AM2107325,book2017}, and a series of promising applications have been achieved. Representative examples include enantio-sensitive controls~\cite{NatPhoton13866}, optofluidic sorting of particles~\cite{NatCom2014}, chiral negative refractive index materials~\cite{Science2004}, and chiral optical devices~\cite{Science2016,PRL114053903}. Chaos describes certain nonlinear dynamics which is sensitive to initial conditions, and has broad applications in physics, chemistry, and engineering \cite{chaosbook}. Thus, the chaos combining with chirality may result in new features for chaos applications. Moreover, existing chiral parameters~\cite{RevModPhys711745} are not suitable for characterizing and quantifying the degree of chirality in chaos.

The radiation-pressure-based hybrid systems are ideal platforms for studying nonlinear physics and chaos~\cite{NatPhoton103992016,PRL1140136012015,PRL111053602}, and can be used for high precision measurements of weak signals~\cite{Braginsky1,Braginsky2}. Advanced micro-/nano-fabrication technologies make the sizes of chaos-based sensors reach the micron level~\cite{RMP861391}, which utilize the abrupt changes of observables in the vicinity of critical points of bifurcation or chaos~\cite{NatureMaterials2011,Appl98,IEEE46440}. Such sensors possess the advantages of high sensitivity and fast response, but their effectiveness is severely limited by the type of detected signals~\cite{NatPhoton9151,Appl98}. For uncontrolled signals exerting on sensors, e.g., optical signals with random phases or mechanical displacements with uncertain directions, even if their amplitudes exceed a critical threshold, abrupt changes (e.g., order-to-chaos, bifurcation cascade) do not necessarily occur~\cite{Nature363411}. Thus, the occurrence of false negative errors during the sensing process is hard to avoid, which hinders the applications of chaos-based sensors~\cite{NatPhoton9151}.

In this Letter, we propose a controllable chiral chaotic device using whispering gallery mode (WGM) resonators as controlled objects and tips as controllers. Chiral chaos is characterized by phase diagrams, Lyapunov exponents, bifurcation diagrams, and other means~\cite{M.C.Gutzwiller}. Underlying physics behind chirality is the regulation of time-reversal symmetry by the photonic gauge potentials generated via tips. We also propose appropriate parameters to quantify chiral chaos. By tuning the chiral order-to-chaos phase transition routes, we construct a type of sensing window that is immune to false negative errors. Through analysis and comparison, we demonstrate the superiority of our sensing proposal in sensing performance compared to existing chaos-based proposals.

\begin{figure}[t]
\centering
\includegraphics[width=8.5cm]{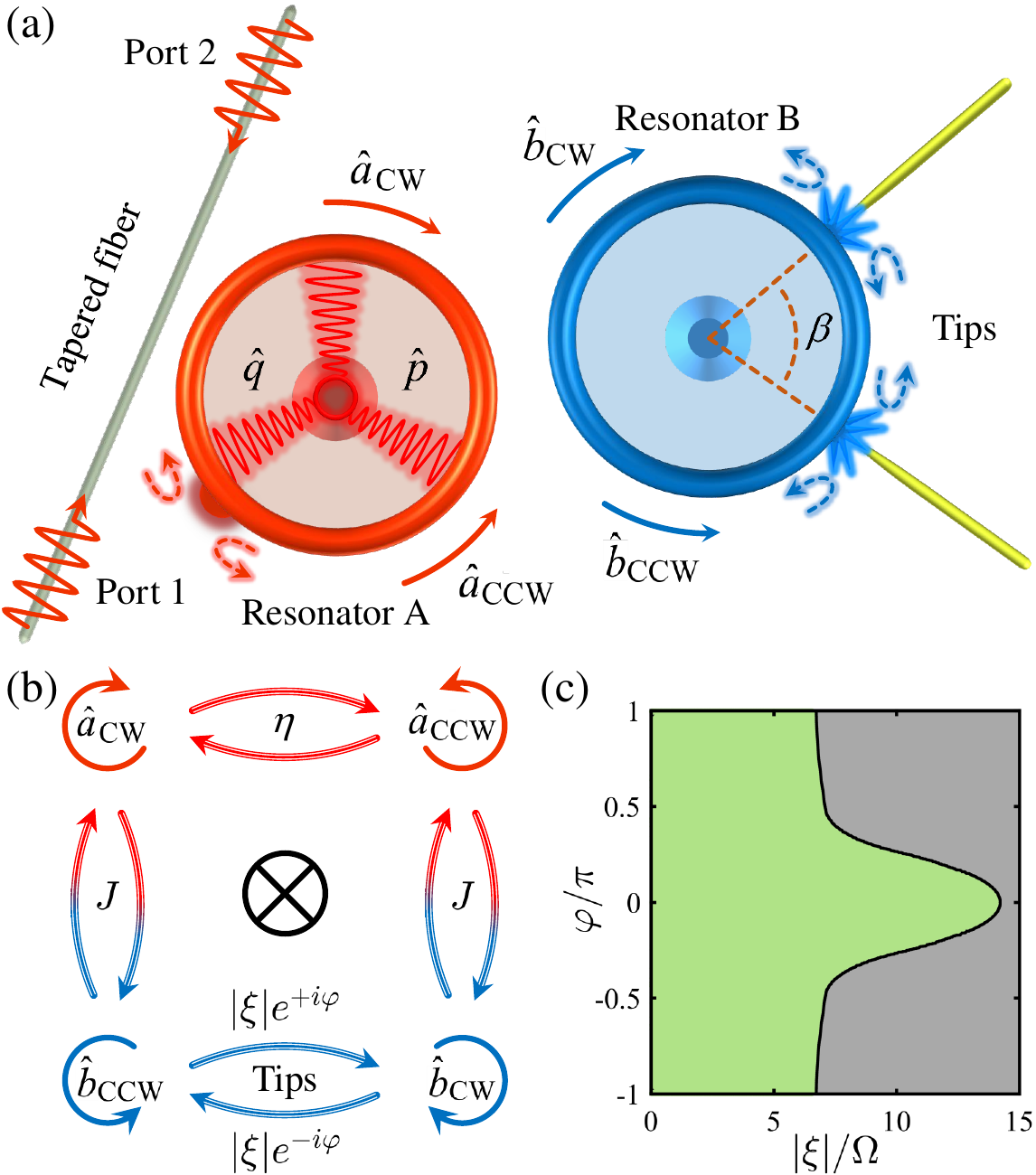}
\caption{(a) Schematic diagram of the proposed chiral chaotic device, where an optomechanical resonator A is coupled to an optical resonator B, and the pumping field is input into resonator A through the tapered fiber from either port. Two tips acting as Rayleigh scatterers contact the rim of resonator B. (b) Loops consisting of optical modes and their couplings. (c) The abilities of tips to regulate $|\xi|$ and $\varphi$ under common experimental conditions (see Supplemental Material \cite{SupMat} for details), where the green and gray regions demarcated by the black curve represent the achievable and unachievable parameter ranges, respectively.}
\label{Fig1}
\vspace{-0.3cm}
\end{figure}

\textit{Chiral chaotic device}.---As schematically illustrated in Fig.~\ref{Fig1}(a), two resonators are coupled via the evanescent fields, and each supports two counterpropagating optical WGMs~\cite{Science2014}. We define $\hat{a}_{\rm CW}$ and $\hat{a}_{\rm CCW}$ ($\hat{b}_{\rm CW}$ and $\hat{b}_{\rm CCW}$) as the annihilation operators for CW and CCW modes in resonator A (B), with a degenerate resonance frequency $\omega_{\rm A}$ $(\omega_{\rm\;\;\!\!\! B})$ and damping rate $\kappa$ $(\gamma)$, respectively. Resonator A supports a mechanical mode described by displacement $\hat{q}$ and momentum $\hat{p}$ operators, with the frequency $\Omega$ and decay rate $\Gamma$. Chaotic dynamics of the device originates from the nonlinear optomechanical interactions between the optical modes and the mechanical mode in resonator A, i.e., $G\;\;\!\!\!(\hat{a}^{\dag}_{\rm CW}\hat{a}_{\rm CW}+\hat{a}^{\dag}_{\rm CCW}\hat{a}_{\rm CCW})\;\;\!\!\!\hat{q}$, where $G$ represents the single-photon optomechanical coupling strength~\cite{RMP861391}.

Couplings between optical modes in two resonators are schematically shown in Fig.~\ref{Fig1}(b) and are described by the effective Hamiltonian ($\hbar=1$)
\begin{align}
\hat{H}_{\rm eff}=\,&-J\hat{a}_{\rm CW}\hat{b}^{\dag}_{\rm CCW}-|\xi|e^{i\varphi}\hat{b}_{\rm CCW}\hat{b}^{\dag}_{\rm CW} \nonumber \\[3pt]
&-J\hat{b}_{\rm CW}\hat{a}^{\dag}_{\rm CCW}-\eta\;\;\!\!\!\hat{a}_{\rm CCW}\hat{a}^{\dag}_{\rm CW}+\rm H.c.,
\label{Eq1}
\end{align}
where $\eta$ is the backscattering strength between CW and CCW modes in resonator A caused by surface defects~\cite{RMP872015}, $|\xi|e^{i\varphi}\;\!\!$ and $|\xi|e^{-i\varphi}\;\!\!$ denote the complex scattering strength between WGMs in resonator B, in which both $|\xi|$ and $\varphi$ can be tuned through tips~\cite{PRL112203901,SupMat}, and $J$ is the coupling strength between two resonators.

\begin{figure}[t]
	\centering
	\includegraphics[width=8.6 cm]{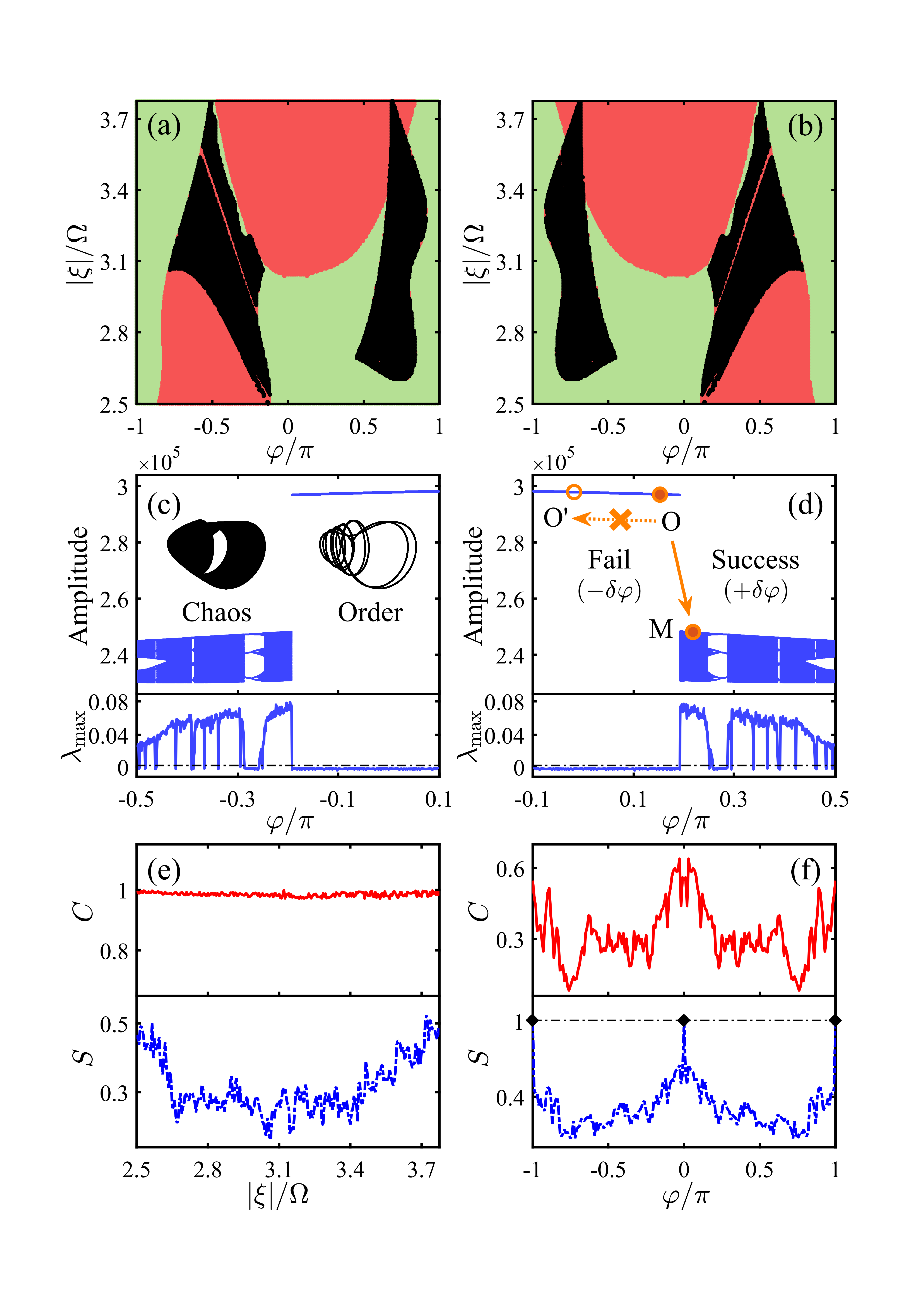}
	\caption{(a),(b) Phase diagrams for different input ports, with (a) and (b) corresponding to input from port 1 and port 2, respectively. The green, red, and black regions indicate self-induced oscillation, period-doubling bifurcations, and chaos, respectively. (c),(d) Bifurcation diagrams of the mechanical oscillation (upper panels) for pumping from port 1 and port 2 and the corresponding maximal Lyapunov exponent spectra (bottom panels), with $|\xi|/\Omega=3$. Dash-dotted lines mark $\lambda_{\rm max}=0$. Insets in (c) present two types of phase-space trajectories of the light fields, i.e., chaos and order. (e),(f) Chirality $C$ and symmetry $S$ between phase diagrams (a) and (b). The black dash-dotted line in (f) marks $S=1$, and the three black diamonds represent $\varphi=0$ and $\pm\;\!\pi$. Parameters~\cite{RMP861391,AOP7}:  $\eta/\Omega=0.15,$ $J/\Omega=2,$ $\Gamma/\Omega= 5\times10^{-3},$ $\gamma/\Omega=5,$ $\kappa/\Omega=0.25,$ $G/\Omega =5\times10^{-5},$ $\varepsilon/\Omega=5.8\times10^{4},$ $(\omega_{\rm A}\;\!\!-\;\;\!\!\!\omega)/\Omega=(\omega_{\rm\;\;\!\!\! B}\;\!\!-\;\;\!\!\!\omega)/\Omega=$ $-\;\!0.5$. }
	\label{Fig2}
	\vspace{-0.3cm}
\end{figure}

\textit{Chiral chaos and its metrics}.---With a monochromatic pumping field with frequency $\omega$ and amplitude $\varepsilon$ applied to either port, the dynamics of the device can be obtained from the Heisenberg equations of motion~\cite{SupMat}. The phase diagrams are presented in Figs.~\ref{Fig2}(a) and \ref{Fig2}(b), where the chaotic and ordered regimes are divided according to the maximal Lyapunov exponent $\lambda_{\rm max}$ (i.e., chaos: $\lambda_{\rm max}>0$; order: $\lambda_{\rm max}<0$) and power spectrum of the mechanical mode~\cite{book2005}. It is obvious that these two patterns are chiral along the $\varphi\;\;\!\!\!$-axis. With the continuous increase of $\varphi$ from $-\;\!\pi$ to $\pi$, the transition process between order state and chaos state for pumping from port 2 is the reverse of that for pumping from port 1. Furthermore, we note that the order-to-chaos transition can be controlled by tuning the parameters $|\xi|$ and $\varphi$, which corresponds to adjusting two tips in experiments, as shown in Fig.~\ref{Fig1}(c). The features depicted in two phase diagrams indicate that the device supports a chiral chaos that varies with $\varphi$. However, no similar chirality occurs on the $|\xi|$-axis. To illustrate chiral chaos more intuitively, we draw the bifurcation diagrams and the maximal Lyapunov exponent spectra in Figs.~\ref{Fig2}(c) and \ref{Fig2}(d). The chirality observed in the structure of two bifurcation diagrams, especially the positions of periodic windows within the chaotic regime, indicates that chiral chaos can be induced by tuning $\varphi$. The spectra of $\lambda_{\rm max}$ also verifies this conclusion.

Physically, the anomalous phenomena in Fig.~\ref{Fig2} imply that time-reversal symmetry of WGMs has been broken. Specifically, chirality results from the couplings between optical modes, as shown in Fig.~\ref{Fig1}(b) and the Hamiltonian $\hat{H}_{\rm eff}$ in Eq.~\eqref{Eq1}. Photons hopping along two reverse loops accumulate opposite net phases, i.e., hopping along loop $\hat{a}_{\rm CCW}\rightarrow\hat{b}_{\rm CW}\rightarrow\hat{b}_{\rm CCW}\rightarrow\hat{a}_{\rm CW}\rightarrow\hat{a}_{\rm CCW}$ and the reverse loop $\;\!\!\hat{a}_{\rm CCW}\;\!\!\rightarrow\;\!\!\hat{a}_{\rm CW}\;\!\!\rightarrow\;\!\!\hat{b}_{\rm CCW}\;\!\!\rightarrow\hat{b}_{\rm CW}\;\!\!\rightarrow\hat{a}_{\rm CCW}\;\!\!$ accumulate phases $-\;\!\varphi$ and $\varphi$, respectively. By combining the gauge transformation and time-reversal transformation~\cite{PRA82043811,SupMat}, we conclude that once $\varphi\neq\mathbb{Z}\;\;\!\!\!\pi$ (here, $\mathbb{Z}$ denotes arbitrary integer), the light fields in the device no longer hold time-reversal symmetry. Similar to previous works~\cite{PRL126123603,PRL122083903}, the net hopping phase marks the presence of a static optical gauge field, which has nontrivial effects on the dynamics of the device. To illustrate this, we define $\delta I_{\rm A}$ as the light intensity difference in resonator A produced by inputting the same pumping field from two different ports. For steady state, it can be expressed as  $\delta I_{\rm A}^{s}=g\;\;\!\!\!(|\eta\;\;\!\!\!+Fe^{i\varphi}|^2-|\eta\;\;\!\!\!+Fe^{-i\varphi}|^2)$~\cite{SupMat}. We can find that for a trivial gauge phase, i.e., $\varphi=\mathbb{Z}\pi$, $\delta I_{\rm A}^{s}=0$, while for a nontrivial gauge phase, the distribution of values of $\delta I_{\rm A}^{s}$ on the $\varphi\;\;\!\!\!$-axis exhibits chirality. Although the analytical expression of $\delta I_{\rm A}$ for chaotic state is unobtainable, the $\varphi$-dependent chirality  for $\delta I_{\rm A}^{s}$ and $\delta I_{\rm A}$  is the same as the chirality of chaos dynamics shown in Figs.~\ref{Fig2}(a) and \ref{Fig2}(b).

To quantitatively characterize the degree of symmetry and chirality between different routes to chaos, we design the following metrics based on the Lyapunov exponents. For any two $N$-element conditional sets $\{x_{i}\}$ and $\{y_{i}\}$, the system will produce equal-length state sets $\{U_{i}\}$ and $\{V_{i}\}$, $i=1,2,\;\;\!\!\!\cdots\;\!\!,N$, respectively. By applying efficient algorithms~\cite{Meccanica1591980,Physicald}, one can calculate two corresponding maximal Lyapunov exponent sets $\{\lambda_{\rm max}^i\}$ and $\{\Lambda_{\rm max}^i\}$. Then we define a class of structural parameters as
\begin{align}	
S=&\;1-\dfrac{1}{N}\sum_{i=1}^{N}\dfrac{|\lambda_{\rm max}^{i}-\Lambda_{\rm max}^{i}|}{|\lambda_{\rm max}^{i}|+|\Lambda_{\rm max}^{i}|}, \label{Eq2} \\[4pt]  C=&\;1-\dfrac{1}{N}\sum_{i=1}^{N}\dfrac{|\lambda_{\rm max}^{i}-\Lambda^{N-i+1}_{\rm max}|}{|\lambda_{\rm max}^{i}|+|\Lambda^{N-i+1}_{\rm max}|}.
\label{Eq3}
\end{align}
It is obvious that $S$ and $C$ reflect the degree of symmetry and chirality between evolution processes $\{U_{i}\}$ and $\{V_{i}\}$. For the ideal chirality and complete achirality, $C$ takes the values 1 and 0, respectively, while for all other cases, $C\;\!\!\in\;\!\!(0,1)$. Similarly, $S$ will reach the value of $1$ for complete symmetry and $0$ for complete asymmetry, while for all other cases, $S\;\!\!\in\;\!\!(0,1)$. More discussions about $S$ and $C$ are provided in Supplemental Material~\cite{SupMat}.

We now demonstrate the effectiveness of metrics $S$ and $C$. As shown in Fig.~\ref{Fig2}(e), the $\varphi\;\;\!\!\!$-dependent chiral order-to-chaos transition process (for each $|\xi|$, $\varphi\;\!\!\in\;\!\![\;\;\!\!\!-\;\!\pi,\pi\;\;\!\!\!]$) shows strong consistency with the value of $C$. Moreover, the relation between $S$ and $|\xi|$ aligns well with the structural characteristics of phase diagrams. In Fig.~\ref{Fig2}(f), the $|\xi|$-dependent order-to-chaos transition process (for each $\varphi$, $|\xi|/\Omega\in\;\!\![\;\;\!\!\!2.5,\;\;\!\!\!3.8\;\;\!\!\!]$) of two phase diagrams in Figs.~\ref{Fig2}(a) and \ref{Fig2}(b) can also be accurately described by $S$-curve and $C$-curve. It is also worth noting that $S$ and $C$ have great distinguishing performance, as shown by the three black diamonds in Fig.~\ref{Fig2}(f), where complete symmetry result in sharp rise of S.

\begin{figure}[t]
\centering
\includegraphics[width= 8.6 cm]{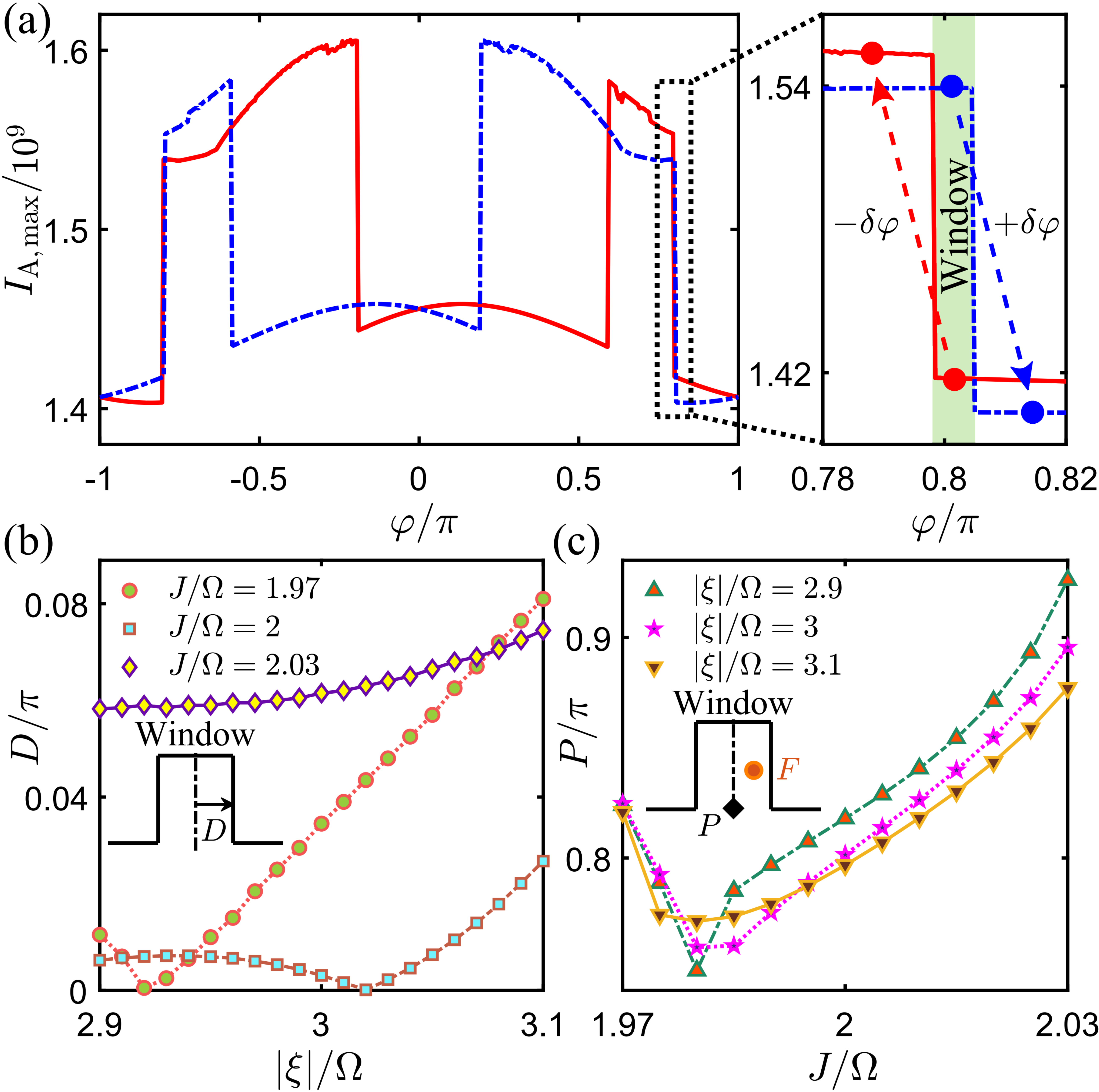}
\caption{(a) The maximum magnitude of the light intensity in resonator A (i.e., $I_{\rm A,\;\! max}=\;\;\!\!\!{\rm max}\;\!\{|a_{\rm CW}(\tau)|^2+\;\;\!\!\!|a_{\rm CCW}(\tau)|^2\}$) versus $\varphi$, with $|\xi|/\Omega=3$. The red and blue curves represent the results of pumping from port 1 and port 2, respectively. The sensing window within the black dotted box is enlarged for convenient observation. The width of the sensing window is determined by the difference of two phase transition points induced by pumping from different ports, i.e., the width of the green strip. (b) and (c) show the adjustability of the window half-width $D$ and center position $P$, respectively. Point $F$ in (c) represents the working point of the device. Other parameters are the same as in Fig.~\ref{Fig2}.}
\label{Fig3}
\vspace{-0.3cm}
\end{figure}

\textit{Chiral chaos for sensitive sensing}.---We observe that a step-like rise or drop in Figs.~\ref{Fig2}(c) and \ref{Fig2}(d) marks the order-to-chaos phase transition, which can be verified by power spectra of the light fields in resonators~\cite{M.C.Gutzwiller}. Near the critical point, a small perturbation $\delta\varphi$ may cause significant changes in the system states, as shown in the insets in Fig.~\ref{Fig2}(c) and the process $\rm O\rightarrow M$ (orange solid arrow) in Fig.~\ref{Fig2}(d). Such a transition has potential applications in high-precision sensing~\cite{NatPhoton9151,IEEE46440}. However, the orientation randomness of the unknown perturbation $\delta\varphi$, i.e., positive deviation $\delta\varphi>0$ (denoted by $+\;\;\!\!\!\delta\varphi)$ or negative deviation $\delta\varphi\;\!\!<\;\!\!0$ (denoted by $-\;\;\!\!\!\delta\varphi)$,  inevitably results in false negative errors for conventional chaos sensing without chirality during the sensing process, as shown in the $\rm O\rightarrow O'\;\!\!$ error (orange dotted arrow) in Fig.~\ref{Fig2}(d). This error seriously limits the detection efficiency.

In our chiral chaotic device, the transitions of order-to-chaos and period-doubling cascades no longer fully overlap for different pumping ports. As shown in Fig.~\ref{Fig3}(a), the distribution of the light intensity from different pumping ports satisfies chiral symmetry. Thus, we can synthesize several stable sensing windows, where the detection of small signal can be realized by splitting it into two which are alternatively switched to two ports. In this case, the dynamics of the device is only sensitive to the strength $|\delta\varphi|$ of the perturbation $\delta\varphi$  no matter what  the deviation orientation of $\delta\varphi$  is.

Performance of the sensing window mainly depends on its half-width $D$, center position $P$, and working point $F$. We now investigate the performance of the right sensing window in Fig.~\ref{Fig3}(a). The selection of the working point $F$ is done according to the detection accuracy and is independent of $D$ and $P$.   However, as  shown in Figs.~\ref{Fig3}(b) and \ref{Fig3}(c),  $D$ and $P$ strongly depend on  the parameters $J$, $|\xi|$ and $\varphi$ of the device. The parameter $J$ can be adjusted via a translation stage~\cite{Nature548192}, $|\xi|$ and $\varphi$ can be tuned by tips as shown in Fig.~\ref{Fig1}(c),  thus the synthesized sensing windows have excellent flexibility. By comparing the effects of $|\xi|$ and $J$ on the window  in Figs.~\ref{Fig3}(b) and \ref{Fig3}(c), we find that $J$ plays dominant role. This indicates that  the variation of the distance between resonators is suitable as a coarse tuning method, while tips are suitable as fine adjustment knobs. Based on the expected accuracy of the detection and chosen value of the half-width $D$, the sensing windows and their applications can be further divided, e.g., narrow windows with small $D$ values are applicable for high precision sensing, wide windows with large $D$ values are applicable for threshold alarms~\cite{J.S.Wilson}. Hereafter, we only discuss the narrow window case.

The above $\varphi$-dependent sensing can be generalized to other sensing parameters. We now study sensing windows constructed on the $(\varepsilon,\omega)$ plane of the pumping field~\cite{SupMat}.  We assume that an additional detected signal with frequency $\omega+\delta\omega$ ($\delta\omega\ll\omega$)  and amplitude $\delta\varepsilon$ ($\delta\varepsilon\ll\varepsilon$)  is applied to the device through the input port. Thus, the total input field is $E_{\rm tot}\;\!\!=\;\;\!\!\!\varepsilon_{\rm tot}\;\;\!\!\!{\rm cos}\;\!\;\!\!(\;\!\;\!\!\omega\;\;\!\!\!\tau+\theta_{\rm tot})$ with
\begin{gather}
\begin{split}
\varepsilon_{\rm tot}&=\;\;\!\!\!\sqrt{\varepsilon^{2}+\delta\varepsilon^{2}+2\;\;\!\!\!\varepsilon\;\;\!\!\!\delta\varepsilon\;\!\!\;\!{\rm cos}\;\;\!\!\!(\delta\omega\;\;\!\!\!\tau+\theta)}\;\;\!\!\!,\\[3pt]
\theta_{\rm tot}&=\;\;\!\!\!\arctan\!\left[\;\;\!\!\!\dfrac{\delta\varepsilon\;\!{\rm sin}\;\;\!\!\!(\delta\omega\;\;\!\!\!\tau+\theta)}{\varepsilon+\delta\varepsilon\;\;\!\!\!{\rm cos}\;\;\!\!\!(\delta\omega\;\;\!\!\!\tau+\theta)}\;\!\!\;\!\right]\!\;\!\!.
\end{split}
\tag{4}\label{Eq4}
\end{gather}
$\varepsilon_{\rm tot}$ and $\theta_{\rm tot}$ are the equivalent amplitude and additional phase, $\theta$ is the initial phase difference between the pumping field and detected signal. To characterize the sensing performance, we classify the sensing results into two categories, i.e., failure and success based on following criteria:  (i)  chiral order-to-chaos phase transition is induced, (ii) the amplitudes of physical quantities (e.g., light intensity and mechanical oscillation amplitude) are significantly changed. If both criteria are met after the detected signal is introduced, then we call it as a success, otherwise as a failure.

\begin{figure}[t]
	\centering
	\includegraphics[width= 8.6 cm]{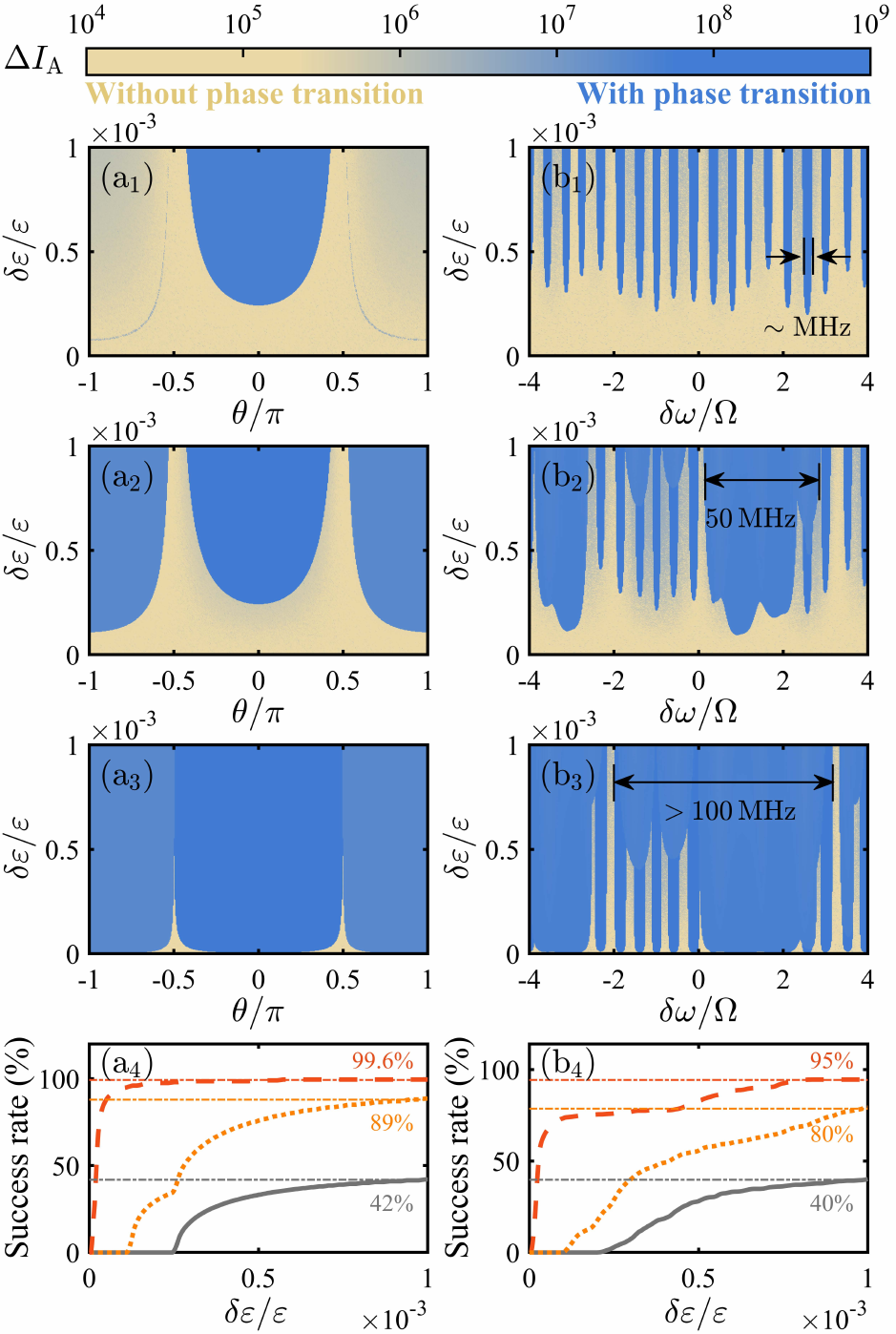}
	\caption{($\rm{a}_{1}$),($\rm{b}_{1}$) Results of conventional chaos-based sensing proposal, obtained by inputting the pumping field and detected signal from port 2. ($\rm{a}_{2}$)($\rm{a}_{3}$) and ($\rm{b}_{2}$) ($\rm{b}_{3}$) Results of chiral-chaos-based sensing proposal using constructed sensing windows. $\Delta I_{\rm A}$ is the change in the maximum light intensity in resonator A before and after the detected signal is introduced. The yellow and blue regions represent sensing failure and success, respectively. We summarize the success rates of ($\rm{a}_{1}$)--($\rm{a}_{3}$) and ($\rm{b}_{1}$)--($\rm{b}_{3}$) in ($\rm{a}_{4}$) and ($\rm{b}_{4}$), respectively, where the gray solid curves, orange dotted curves, and red dashed curves correspond to ($\rm{a}_{1}$)($\rm{b}_{1}$), ($\rm{a}_{2}$)($\rm{b}_{2}$), and ($\rm{a}_{3}$)($\rm{b}_{3}$), respectively. The success rate here is defined as the proportion of successful sensing events to the total events, for weak signals with a fixed amplitude $\delta\varepsilon$. Parameters: ($\rm{a}_{1}$)--($\rm{a}_{4}$) $\delta\omega=0$, ($\rm{b}_{1}$)--($\rm{b}_{4}$) $\theta/\pi=0.5$, ($\rm{a}_{1}$)($\rm{a}_{2}$) and ($\rm{b}_{1}$)($\rm{b}_{2}$) $\varphi/\pi=0.505$, ($\rm{a}_{3}$) and ($\rm{b}_{3}$) $\varphi/\pi=0.5$. For more details, see Supplemental Material \cite{SupMat}.}
	\label{Fig4}
	\vspace{-0.3cm}
\end{figure}

We first consider the special case that the detected signal is in resonance with the pumping field, i.e., $\delta\omega=0$, and the results are shown in Figs.~\ref{Fig4}($\rm{a}_{1}$)--($\rm{a}_{4}$). For a conventional sensing using order-to-chaos phase transition~\cite{IEEE46440}, as shown in  Fig.~\ref{Fig4}($\rm{a}_{1}$), its performance is strongly affected by the value of $\theta$, however $\theta$ cannot be known a priori in the actual sensing process, which will inevitably lead to failure in some cases. Thus, the success rate is below 50\% as shown in the gray solid curve in Fig.~\ref{Fig4}($\rm{a}_{4}$).  In contrast to conventional one, the chiral chaos based sensing  exhibits stronger robustness to $\theta$ as shown in Fig.~\ref{Fig4}($\rm{a}_{2}$) for given $\varphi$,  e.g., $\varphi/\pi=0.505$, and the corresponding success rate is also significantly enhanced as shown in the orange dotted curve in Fig.~\ref{Fig4}($\rm{a}_{4}$). By further optimizing $\varphi$, we can even achieve sensing that is immune to $\theta$, and the success rate can approach 100\%, as shown in Fig.~\ref{Fig4}($\rm{a}_{3}$) and the red dashed curve in Fig.~\ref{Fig4}($\rm{a}_{4}$)  for,  e.g., $\varphi/\pi=0.5$.

For $\delta\omega\neq0$,  the effective detectable range of $\delta\omega$, defined as continuously range of $\delta\omega$ for sensing success, is an important index to evaluate the quality of a sensor. Fig.~\ref{Fig4}($\rm{b}_{1}$) shows a serious flaw of conventional proposal, i.e., effective detectable range is very narrow, only a few $\rm MHz$. However, our proposal utilizing chiral chaos has great advantages in this aspect, as shown in Fig.~\ref{Fig4}($\rm{b}_{2}$) for given $\varphi$, e.g., $\varphi/\pi=0.505$, an  effective detectable range with tens of $\rm MHz$ is achieved. By modulating the chiral chaos by optimizing $\varphi$,  the effective detectable range can exceed 100 $\rm MHz$ as shown in Fig.~\ref{Fig4}($\rm{b}_{3}$). Besides, the success rates in Fig.~\ref{Fig4}($\rm{b}_{4}$) also prove the superiority of our proposal. In practice, the effective detectable range is also affected by the bandwidth of the pumping field.

We further analyze these results. As shown in Eq.~\eqref{Eq4}, if $\delta\omega=0$,  $\theta$ controls the enhancement ($\varepsilon_{\rm tot}>\varepsilon$)  or attenuation ($\varepsilon_{\rm tot}<\varepsilon$)  of the pumping field. This implies that the upper limit of the success rate is only 50\% for conventional proposal, since the order-chaos phase transition can only follow one route no matter how the signal is applied. However, the situation is changed through chiral chaos. By splitting the signal into two which are alternatively switched to two ports, order-chaos phase transition through a sensing window can proceed along two opposite routes~\cite{SupMat}, which ensure that the theoretical upper limit of the success rate is 100\%.

For $\delta\omega\neq0$, $\theta$ mainly affects the initial conditions. The modulation of the pumping field by the detected signal is nonmonotonic, $\varepsilon_{\rm tot}$ is periodically less or more than $\varepsilon$, and $\theta_{\rm tot}$ also varies periodically. The system undergoes chaos-to-order transition within half the time, while the remaining half is dedicated to maintaining the current state. The final effect of the competition process depends on the skeleton of the chaotic attractors and the initial conditions~\cite{book2004}. For chiral chaos, the structures of attractors corresponding to different input directions are inconsistent, resulting in more attractors gathered in the region of the sensing window than those of conventional chaos. Thus, the effective detection range for our chiral-chaos-based sensing is wider than that of convention one.

\textit{Discussion and conclusion}.---In summary, we propose chiral chaos and construct a chiral chaotic device. The phase diagrams of chiral chaos and chaotic dynamics are carefully investigated. Moreover, we introduce two structure parameters as metrics of symmetry and chirality. Based on chiral chaos,  we show that the false negative errors plaguing conventional chaos-based sensing proposals~\cite{IEEE46440,NatPhoton9151} can be partially or even completely overcome.  
In our study for chiral-chaos-based sensing, the required quality factor $Q\in[10^{7},10^{8}]$ for the WGM resonators is achievable~\cite{NatPhoton4462010,PRL104083901}.  The controllable coupling parameters can be realized via the translation stages~\cite{OE1823535,AOP7}. Moreover, these parameters can be further optimized by changing the medium environment or electroplating specific materials on tips~\cite{Nature548192}. We finally mention that our proposal can also be generalized to other physical systems, e.g., superconducting quantum circuits~\cite{PhysRep.718.1} and systems of quantum acoustics~\cite{Shao-yang}.

\textit{Acknowledgments}.---Y.X.L. is supported by the National  Natural Science Foundation of China with Grants No.~12374483 and No.~92365209. J.Z. is supported by Tsinghua-Foshan Innovation Special Fund (20182000306); Department of Liaoning Province and State Key Laboratory of Robotics, China (2021-KF-22-01); Wenhai Program of the S\&T Fund of Shandong Province for Pilot National Laboratory for Marine Science and Technology (Qingdao) (NO.2021WHZZB2500, NO.2021WHZZB2501).

\clearpage
\onecolumngrid
\flushbottom
\begin{center}
	\textbf{\large Supplemental Material: Chiral Chaos Enhanced Sensing}
\end{center}
\setcounter{equation}{0}
\setcounter{figure}{0}
\renewcommand{\thefigure}{S\arabic{figure}}
\renewcommand{\theequation}{S\arabic{equation}}
\setcounter{secnumdepth}{3}
\makeatletter
\def\@hangfrom@section#1#2#3{\@hangfrom{#1#2#3}}
\makeatother

\section{Details of the chiral chaotic device presented in the main text}

In a rotating frame at the pumping field frequency $\omega$, the Hamiltonian of our system can be written as
\begin{align}
	\hat{H}=&\ \hbar\;\;\!\!\!\Delta_{\rm A}(\hat{a}^{\dag}_{\rm CW}\hat{a}_{\rm CW}+\hat{a}^{\dag}_{\rm CCW}\hat{a}_{\rm CCW})+\frac{\hbar}{2}\;\;\!\!\!\Omega\;\;\!\!\!(\;\;\!\!\!\hat{p}^2+\hat{q}^2) -\hbar \;\;\!\!\!G(\hat{a}^{\dag}_{\rm CW}\hat{a}_{\rm CW}+\hat{a}^{\dag}_{\rm CCW}\hat{a}_{\rm CCW})\hat{q} +\hbar\;\!\!\!\;\Delta_{\rm B}(\hat{b}^{\dag}_{\rm CW}\hat{b}_{\rm CW}+\hat{b}^{\dag}_{\rm CCW}\hat{b}_{\rm CCW}) \nonumber \\
	&-\hbar\;\;\!\!\! J(\hat{a}^{\dag}_{\rm CW}\hat{b}_{\rm CCW}+\hat{a}^{\dag}_{\rm CCW}\hat{b}_{\rm CW}+\hat{a}_{\rm CW}\hat{b}^{\dag}_{\rm CCW}+\hat{a}_{\rm CCW}\hat{b}^{\dag}_{\rm CW})-\hbar\;\;\!\!\!\eta\;\;\!\!\!(\hat{a}^{\dag}_{\rm CW}\hat{a}_{\rm CCW}+\hat{a}_{\rm CW}\hat{a}^{\dag}_{\rm CCW}) \nonumber \\[3pt]
	&-\hbar\;\;\!\!\!|\xi|(e^{i\varphi}\hat{b}^{\dag}_{\rm CW}\hat{b}_{\rm CCW}+e^{-i\varphi}\hat{b}_{\rm CW}\hat{b}^{\dag}_{\rm CCW}) +i\hbar\;\;\!\!\!\varepsilon_{1}(\hat{a}_{\rm CW}^{\dag}-\hat{a}_{\rm CW})+i\hbar\;\;\!\!\!\varepsilon_{2}(\hat{a}_{\rm CCW}^{\dag}-\hat{a}_{\rm CCW}).
\end{align}	
Dynamics of the above hybrid optomechanical system are described by the Heisenberg equations of motion
\begin{align}
	\dot{a}_{\rm CCW}&=-i(\Delta_{\rm A}-i\kappa)a_{\rm CCW}+iGa_{\rm CCW}q+i\eta\;\;\!\!\! a_{\rm CW}+iJb_{\rm CW}+\varepsilon_{2},\label{S2}\\[3pt]
	\dot{a}_{\rm CW}&=-i(\Delta_{\rm A}-i\kappa)a_{\rm CW}+iGa_{\rm CW}q+i\eta\;\;\!\!\! a_{\rm CCW}+iJb_{\rm CCW}+\varepsilon_{1},\label{S3}\\[3pt]
	\dot{b}_{\rm CCW}&=-i(\Delta_{\rm B}-i\gamma)b_{\rm CCW}+i|\xi|e^{-i\varphi}b_{\rm CW}+iJa_{\rm CW},\label{S4}\\[3pt]			
	\dot{b}_{\rm CW}&=-i(\Delta_{\rm B}-i\gamma)b_{\rm CW}+i|\xi|e^{i\varphi}b_{\rm CCW}+iJa_{\rm CCW},\label{S5}\\[3pt]	
    \dot{q}&=\Omega\:\! p,\quad \dot{p}=-\;\!\Omega\:\! q+G(|a_{\rm CW}|^{2}+|a_{\rm CCW}|^{2})-\Gamma\:\! p,\label{S6}
\end{align}
where $a_{\rm CW}=\langle\hat{a}_{\rm CW}\rangle$ and $a_{\rm CCW}=\langle\hat{a}_{\rm CCW}\rangle$ ($b_{\rm CW}=\langle\hat{b}_{\rm CW}\rangle$ and $b_{\rm CCW}=\langle\hat{b}_{\rm CCW}\rangle$) represent the amplitudes of CW and CCW modes in resonator A (B), $q=\langle\hat{q}\;\;\!\!\!\rangle$ and $p=\langle\;\;\!\!\!\hat{p}\;\;\!\!\!\rangle$ are the displacement and momentum of the mechanical mode (for any operator $\hat{o}$, $o\equiv\langle\hat{o}\;\;\!\!\!\rangle$ denotes its expectation value). Here, correlations between optical modes and mechanical mode have been ignored~\cite{PRL111073603,PRL103213603}. $\Delta_{\rm A}=\;\;\!\!\!\omega_{\rm A}\;\!\!-\omega\ (\Delta_{\rm B}=\;\;\!\!\!\omega_{\rm \;\;\!\!\!B}-\omega)$ denotes the detuning between WGMs in resonator A (B) and the pumping field, $\kappa$ $(\gamma\;\;\!\!\!)$ is the damping rate of WGMs in resonator A (B), $\varepsilon_{j}$ denotes the amplitude of the pumping field at port $j$ ($\;\;\!\!\!j=1,2$), $\Omega$ and $\Gamma$ are the mechanical resonance frequency and damping rate, $G$ is the single-photon optomechanical coupling strength, $\eta$ and $|\xi|e^{\pm i\varphi}$ are the backscattering strengths of WGMs in resonators A and B, respectively, and $J$ is the tunneling strength between resonators. We now provide more details on it.

\subsection{Tips-induced coupling between CW and CCW modes under the dipole approximation}

The electric field distribution in resonator B is
\begin{gather}
		\hat{\boldsymbol{E}}(\boldsymbol{r},t)=f(\boldsymbol{r})\sqrt{\frac{\hbar\;\!\omega_{\rm B}}{2\;\;\!\!\!\varepsilon_{0}V_{\rm B}}}\;\;\!\!\!(\hat{b}_{\rm CW}e^{-i\;\;\!\!\!\omega_{\rm B}t+i\boldsymbol{k}\cdot\boldsymbol{r}}+\hat{b}_{\rm CCW}e^{-i\;\;\!\!\!\omega_{\rm B}t-i\boldsymbol{k}\cdot\boldsymbol{r}}+{\rm H.c.})\;\!\;\!\!\boldsymbol{e}(\boldsymbol{r}),
\end{gather}
where $\omega_{\rm\;\;\!\!\! B}$ and $V_{\rm B}$ denote the resonant frequency and quantization volume of WGMs in resonator B, $\varepsilon_{0}$ is the vacuum permittivity, $\hat{b}_{\rm CW}$ and $\hat{b}^{\dagger}_{\rm CW}$ ($\hat{b}_{\rm CCW}$ and $\hat{b}^{\dagger}_{\rm CCW}$) represent the annihilation and creation operators of CW (CCW) mode. Here, $\boldsymbol{k}$ $\;\!\!(-\boldsymbol{k})\;\!\!$ denotes the wave vector of CW $\!\!\;$(CCW)$\;\!\!$ traveling wave mode, $\boldsymbol{e}(\boldsymbol{r})\;\!\!$ and $\!f(\boldsymbol{r})\;\!\!$ represent the unit vector of electric field direction and cavity-mode function at position $\boldsymbol{r}$. Two tips contact the rim of resonator B at $\boldsymbol{r}_{1}$ and $\boldsymbol{r}_{2},$ causing them to be polarized by the evanescent fields, which can be described as
\begin{gather}
	\hat{\boldsymbol{P}}(\boldsymbol{r}_{1},t)=\varepsilon_{0}\;\;\!\!\!\chi_{1}\hat{\boldsymbol{E}}(\boldsymbol{r}_{1},t),\quad \hat{\boldsymbol{P}}(\boldsymbol{r}_{2},t)=\varepsilon_{0}\;\;\!\!\!\chi_{2}\;\;\!\!\!\hat{\boldsymbol{E}}(\boldsymbol{r}_{2},t),
\end{gather}
in which $\chi_{1}$ and $\chi_{2}$ represent the polarizability of tip $1$ and tip $2$. Utilizing the dipole approximation and rotating wave approximation, and considering the backscattering process with strength $\xi_{0}$ caused by surface defects, the interaction energy between CW and CCW modes can be obtained as~\cite{PRA840638282011}
\begin{align}
		\hat{V}=&-\frac{1}{2}\;\;\!\!\!\hat{\boldsymbol{P}}(\boldsymbol{r}_{1},t)\;\!\!\cdot\!\hat{\boldsymbol{E}}(\boldsymbol{r}_{1},t)-\frac{1}{2}\;\;\!\!\!\hat{\boldsymbol{P}}(\boldsymbol{r}_{2},t)\;\!\!\cdot\!\hat{\boldsymbol{E}}(\boldsymbol{r}_{2},t)-\hbar\;\!\xi_{0}(\hat{b}_{\rm CW}^{\dag}\hat{b}_{\rm CCW}+\hat{b}_{\rm CW}\hat{b}^{\dag}_{\rm CCW}) \nonumber \\
		=&-\hbar\;\!\delta(\hat{b}_{\rm CW}^{\dag}\hat{b}_{\rm CW}+\hat{b}^{\dag}_{\rm CCW}\hat{b}_{\rm CCW})-\hbar\;\;\!\!\!|\xi|(e^{i\varphi}\hat{b}^{\dag}_{\rm CW}\hat{b}_{\rm CCW}+e^{-i\varphi}\hat{b}_{\rm CW}\hat{b}^{\dag}_{\rm CCW}),
\end{align}
in which
\begin{gather}
   \delta=\frac{\omega_{\rm B}(\chi_{1}f^{2}(\boldsymbol{r}_{1})+\chi_{2}f^{2}(\boldsymbol{r}_{2}))}{2V_{\rm B}},
   \quad |\xi|e^{\pm i\varphi}= \xi_{0}+\frac{\omega_{\rm B}(\chi_{1}f^{2}(\boldsymbol{r}_{1})+\chi_{2}f^{2}(\boldsymbol{r}_{2})e^{\pm i2n\beta})}{2V_{\rm B}}.
   \label{S10}
\end{gather}
Note that the dipole-dipole interaction has been ignored owing to the sufficient distance between tips. Without loss of generality, we have chosen $\boldsymbol{r}_{1}$ as the origin of coordinates. $n$ is the azimuthal order corresponding to the wave vector $\boldsymbol{k}$, i.e., the number of wavelengths in the perimeter of resonator B, $\beta$ is the spatial phase difference between two tips, and $\delta$ is the frequency shift of the optical modes induced by two tips, according to the common experiment conditions \cite{OE182010,NatPhoton42010}, we calculate its effective range in Fig.~\ref{Sfigure1}(a).

\subsection{Tips-induced broadening of CW and CCW modes under the Weisskopf-Wigner approximation}

\begin{figure*}[t]
	\centering
	\includegraphics[width= 16 cm]{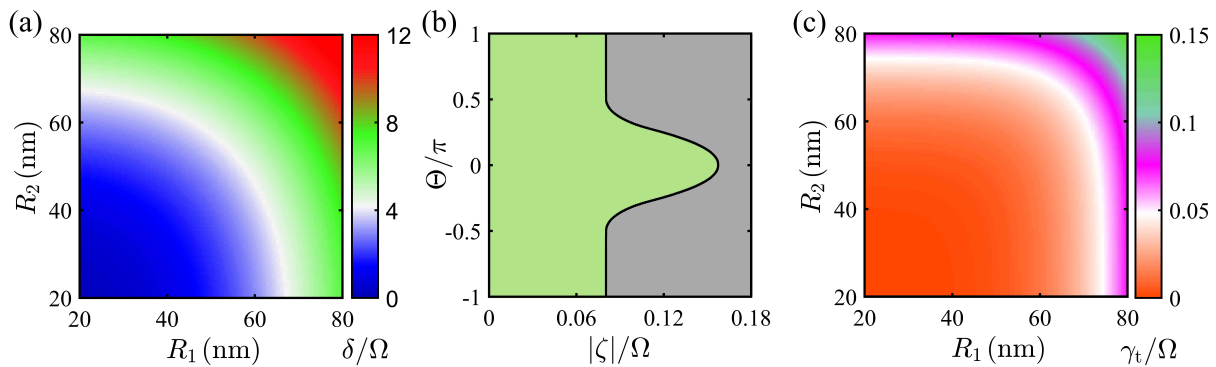}
	\caption{(a) $\delta$ as a function of the geometric scale (effective radius) of two tips. (b) $|\zeta|e^{\pm i\Theta}$ as a function of the geometric scale and relative position of two tips, where the green and gray regions represent the available and unavailable ranges. (c) $\gamma_{\rm t}$ as a function of the geometric scale (effective radius) of two tips. In this Letter, the following parameters are used in all figures, $n_{1}^2=n_{2}^2=3.9$, $\Omega=2\pi\times20$ $\rm MHz$, $f(\boldsymbol{r}_{1})=f(\boldsymbol{r}_{2})=0.3$, $V_{\rm B}=200\;\rm\mu m^{3}$, $v=3\times 10^8\;\rm m/s$, $\omega_{\rm \;\;\!\!\! B}=2\pi \times190\;\rm THz$, $\beta\in[\;\;\!\!\!-\;\!\pi,\pi\;\;\!\!\!]$,   $R_{1}$ and $R_{2}\in [\;\;\!\!\!20,80\;\;\!\!\!]\;\rm nm$.}
	\label{Sfigure1}
	\vspace{-0.3cm}
\end{figure*}

In addition to inducing coupling between optical modes, tips also lead to additional linewidth broadening. The Hamiltonian of the WGMs coupled to the reservoir modes can be expressed as
\begin{gather}
		\hat{H}=\;\;\!\!\!\hbar\;\!\omega_{\rm \;\;\!\!\!B}(\hat{b}^{\dag}_{\rm CW}\hat{b}_{\rm CW}+\hat{b}^{\dag}_{\rm CCW}\hat{b}_{\rm CCW})+\sum_{s}\hbar\;\!\omega_{s}\hat{c}^{\dag}_{s}\hat{c}_{s}-\sum_{l}\sum_{s}\hbar\;\!\;\!\!g_{l,s}(\hat{b}_{\rm CW}^{\dag}\hat{c}_{s}e^{i(\boldsymbol{k}_{s}-\boldsymbol{k})\cdot \boldsymbol{r}_{l}}+\hat{b}_{\rm CCW}^{\dag}\hat{c}_{s}e^{i(\boldsymbol{k}_{s}+\boldsymbol{k})\cdot \boldsymbol{r}_{l}}+\rm H.c.),
\end{gather}
where $\hat{c}_{s}$ and $\hat{c}^{\dag}_{s}$ are the annihilation and creation operators of the $s\rm th$ reservoir mode with frequency $\omega_{s}$, respectively. $g_{l,s}=\chi_{l}f(\boldsymbol{r}_{l})\sqrt{\omega_{\rm \;\;\!\!\!B}\;\!\omega_{s}/4V_{\rm B}V_{\rm vac}}\;\!\boldsymbol{e}_{l}\!\cdot\!\boldsymbol{e}_{s}$ denotes the coupling strength between the optical modes and the $s\rm th$ reservoir mode induced by tip $l$ positioned at $\boldsymbol{r}_{l}$, where $\boldsymbol{e}_{s}$ and $V_{\rm vac}$ denote
the unit vector and quantization volume of the $s\rm th$ reservoir mode, and $\boldsymbol{e}_{l}$ denotes the unit vector of the WGMs. Then the Heisenberg-Langevin equations become
\begin{gather}
		\dot{\hat{b}}_{\rm CW}=-i\;\;\!\!\!\omega_{\rm\;\;\!\!\! B}\hat{b}_{\rm CW}+i\sum_{l}\sum_{s}g_{l,s}\;\;\!\!\!\hat{c}_{s}e^{i(\boldsymbol{k}_{s}-\boldsymbol{k})\cdot \boldsymbol{r}_{l}}, \label{S12} \\
		\dot{\hat{b}}_{\rm CCW}=-i\;\;\!\!\!\omega_{\rm \;\;\!\!\!B}\hat{b}_{\rm CCW}+i\sum_{l}\sum_{s}g_{l,s}\;\;\!\!\!\hat{c}_{s}e^{i(\boldsymbol{k}_{s}+\boldsymbol{k})\cdot \boldsymbol{r}_{l}},  \label{S13} \\
		 \dot{\hat{c}}_{s}=-i\;\;\!\!\!\omega_{s}\hat{c}_{s}+i\sum_{l}g_{l,s}(\hat{b}_{\rm CW}e^{-i(\boldsymbol{k}_{s}-\boldsymbol{k})\cdot \boldsymbol{r}_{l}}+\hat{b}_{\rm CCW}e^{-i(\boldsymbol{k}_{s}+\boldsymbol{k})\cdot \boldsymbol{r}_{l}}\;\!\!).
		 \label{S14}
\end{gather}
Further processing Eq.~(\ref{S14}) by formal integration gives
\begin{align}
\hat{c}_{s}(t)= \hat{c}_{s}(0)e^{-i\;\;\!\!\!\omega_{s}t}+i\sum_{l}g_{l,s}\!\int_{0}^{t}\!\big[\hat{b}_{\rm CW}(t')e^{-i(\boldsymbol{k}_{s}-\boldsymbol{k})\cdot \boldsymbol{r}_{l}}+\hat{b}_{\rm CCW}(t')e^{-i(\boldsymbol{k}_{s}+\boldsymbol{k})\cdot \boldsymbol{r}_{l}}\big]e^{-i\;\;\!\!\!\omega_{s}(t-t')}dt'.
\end{align}
We now separate the slowly varying and fast oscillatory contributions of CW and CCW modes
\begin{align}
	\hat{b}_{\rm CW}(t)=\tilde{b}_{\rm CW}(t)e^{-i\;\;\!\!\!\omega_{\rm B}t}, \quad 	\hat{b}_{\rm CCW}(t)=\tilde{b}_{\rm CCW}(t)e^{-i\;\;\!\!\!\omega_{\rm B}t}.
\end{align}
Then Eqs.~(\ref{S12}) and (\ref{S13}) can be rewritten as
\begin{gather}
\dot{\tilde{b}}_{\rm CW}(t)=-\sum_{l}\sum_{s}g_{l,s}^{2}\!\int_{0}^{t}\!\big[\tilde{b}_{\rm CW}(t')+\tilde{b}_{\rm CCW}(t')e^{-i2\boldsymbol{k}\cdot\boldsymbol{r}_{l}}\big]e^{-i(\omega_{s}-\omega_{\rm B})(t-t')}dt'+i\sum_{l}\sum_{s}g_{l,s}\;\;\!\!\!\hat{c}_{s}(0)e^{-i(\omega_{s}-\omega_{\rm B})t}e^{i(\boldsymbol{k}_{s}-\boldsymbol{k})\cdot \boldsymbol{r}_{l}},\label{S17} \\
\dot{\tilde{b}}_{\rm CCW}(t)=-\sum_{l}\sum_{s}g_{l,s}^{2}\!\int_{0}^{t}\!\big[\tilde{b}_{\rm CCW}(t')+\tilde{b}_{\rm CW}(t')e^{i2\boldsymbol{k}\cdot\boldsymbol{r}_{l}}\big]e^{-i(\omega_{s}-\omega_{\rm B})(t-t')}dt'+i\sum_{l}\sum_{s}g_{l,s}\;\;\!\!\!\hat{c}_{s}(0)e^{-i(\omega_{s}-\omega_{\rm B})t}e^{i(\boldsymbol{k}_{s}+\boldsymbol{k})\cdot \boldsymbol{r}_{l}},\label{S18}
\end{gather}
where the coupling terms between the CW and CCW modes at different positions caused by the reservoir have been neglected. The fluctuation operators associated with $\hat{c}_{s}(0)$ will be ignored in the following, since they have no effect on the mean value equations ($\langle\hat{c}_{s}(0)\rangle=0$). If we further assume that the reservoir modes are closely spaced in frequency, then the following substitution relation holds
\begin{gather}
	\sum_{s}\;\rightarrow\;\frac{2V_{\rm vac}}{(2\pi v)^3}\int_{0}^{2\pi}d\phi\int_{0}^{\pi}{\rm sin}(\theta)d\theta\int_{0}^{\infty}\omega_{s}^{2}d\omega_{s},\label{S19}
\end{gather}
in which we have set the optical fields to be polarized along the $z$-axis, $\theta$ and $\phi$ describe the polarization direction of the reservoir field in spherical coordinate system, and $v$ accounts for the velocity of light in the medium surrounding resonator B. Combining Eqs.~(\ref{S17})--(\ref{S19}), the tips-induced broadening of CW and CCW modes can be obtained as
\begin{gather}
	\dot{\tilde{b}}_{\rm CW}(t)=-\;\!\gamma_{\rm t}\tilde{b}_{\rm CW}(t)-|\zeta|e^{i\Theta}\tilde{b}_{\rm CCW}(t),\quad		
	\dot{\tilde{b}}_{\rm CCW}(t)=-\;\!\gamma_{\rm t}\tilde{b}_{\rm CCW}(t)-|\zeta|e^{-i\Theta}\tilde{b}_{\rm CW}(t),
\end{gather}
in which
\begin{gather}
 \gamma_{\rm t}=\frac{\omega_{\rm B}^{4}(\chi_{1}^{2}f^{2}(\boldsymbol{r}_{1})+\chi_{2}^{2}f^{2}(\boldsymbol{r}_{2}))}{12\pi v^{3}V_{\rm B}},\quad
 |\zeta|e^{\pm i\Theta}=\frac{\omega_{\rm B}^{4}(\chi_{1}^{2}f^{2}(\boldsymbol{r}_{1})+\chi_{2}^{2}f^{2}(\boldsymbol{r}_{2})e^{\pm i2n\beta})}{12\pi v^{3}V_{\rm B}}.
 \label{S21}
\end{gather}
If we further ignore the shape details of tips and treat them as spheres, then the polarizability of tips can be expressed as $\chi_{l}=4\;\;\!\!\!\pi R_{l}^{\;\;\!\!\!3}(n_{l}^2-1)/(n_{l}^2+2)$, where $R_{l}$ and $n_{l}$ denote the effective radius and refractive index of the sphere scatterer equivalent by tip $l$. After combining specific parameters, we calculate the value ranges of $|\zeta|e^{\pm i\Theta}$ and $\gamma_{\rm t}$, as shown in Figs.~\ref{Sfigure1}(b) and~\ref{Sfigure1}(c). Furthermore, Eqs.~(\ref{S10}) and~(\ref{S21}) also state that $\delta$ and $|\xi|e^{\pm i\varphi}$ are proportional to $R_{l}^{\;\;\!\!\!3}$, while $\gamma_{\rm t}$ and $|\zeta|e^{\pm i\Theta}$ are proportional to $R_{l}^{\;\;\!\!\!6}$, thus, for small scale tips $\{|\xi|e^{\pm i\varphi},\delta\}\gg\{|\zeta|e^{\pm i\Theta},\gamma_{\rm t}\}$.

\section{Details of chiral chaos and its metrics}

We numerically solve the equations of motion Eqs.~(\ref{S2})--(\ref{S6}) to obtain the dynamics of the hybrid optomechanical system. To improve the accuracy of calculations and simplify the discussions, we normalize all frequency scales to the mechanical oscillation frequency $\Omega$ and extend the time scale to $\tau = \Omega\;\;\!\!\!t$. Furthermore, we discard the initial transient state $\tau_{0}/2\pi \in [\;\;\!\!\!0, 8000\!\!\!\;\;]$ in all figures. It should be noted that linear and nonlinear interaction terms introduce enormous complexity to the equations of motion, which may lead to the coexistence of different limit cycles in certain regions of parameter space~\cite{PRB1983}. In order to reduce its interference to subsequent discussions, the initial conditions we adopted in all numerical calculations are $a_{\rm CW}(0)=a_{\rm CCW}(0)=b_{\rm CW}(0)=b_{\rm CCW}(0)=q(0)=p(0)=0$.

\subsection{Main operating modes of the chiral chaotic device}

\begin{figure*}[t]
	\centering
	\includegraphics[width= 16 cm]{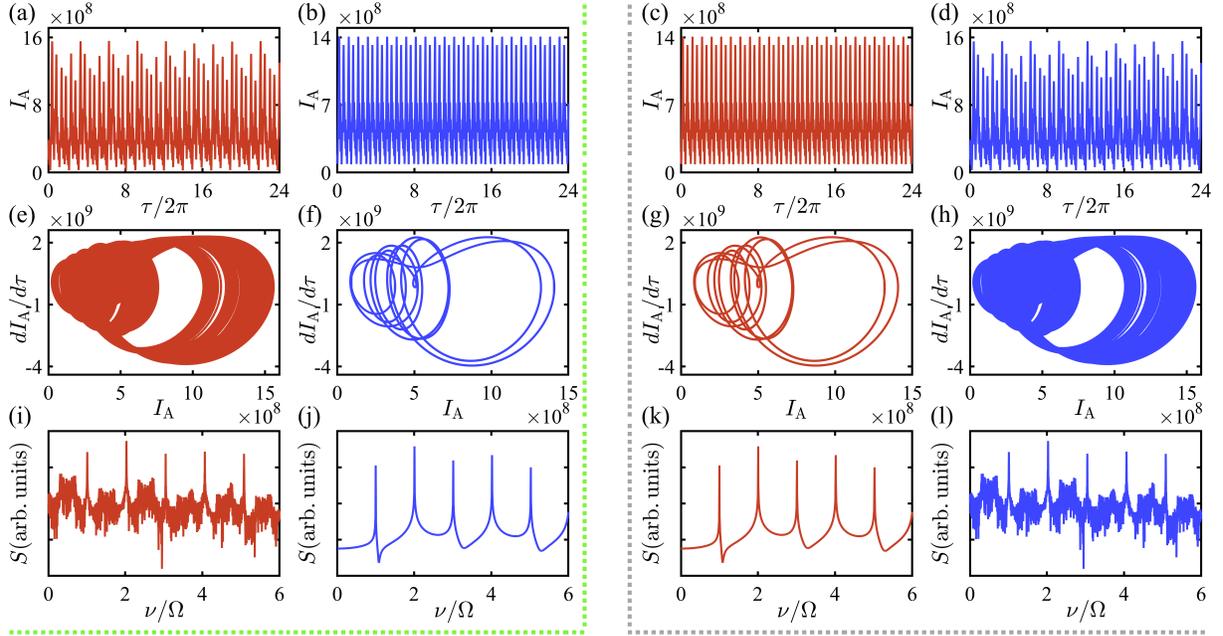}
	\caption{(a)--(d) Time-domain evolutions of the total light intensity in resonator A $(I_{\rm A}=|a_{\rm CW}|^2+|a_{\rm CCW}|^2)$ for pumping from port 1 (red) and port 2 (blue), under different parameter conditions. Here, the hopping phases in the two regions demarcated by the green and gray dotted lines are $\varphi/\pi=0.755$ and $\varphi/\pi=-\;\!0.755$, respectively. (e)--(h) The corresponding phase trajectories. (i)--(l) The corresponding frequency spectra.  Considering the current fabrication capabilities of optical microresonators~\cite{NatPhoton10399,RevModPhys861391}, other parameters chosen here are $\Omega = 2\pi\times20\; \rm MHz$, $G/\Omega = 5\times10^{-5},$ $\Gamma/\Omega = 5\times10^{-3},$ $\Delta_{\rm A}/\Omega=-\;\! 0.5,$ $\Delta_{\rm B}/\Omega=-\;\! 0.5$, $\eta/\Omega=0.15$, $J/\Omega =2$, $|\xi|/\Omega = 3.29$, $\kappa/\Omega=0.25$, $\gamma/\Omega=5$, and $\varepsilon/\Omega = 5.8\times10^{\;\;\!\!\!4}$.}
	\label{Sfigure2}
	\vspace{-0.3cm}
\end{figure*}

Experimentally, the operating mode of our device can be detected by the temporal evolution, power spectrum, and phase diagram of the intracavity light fields~\cite{PRL98167203}. Therefore, we present them all in Fig.~\ref{Sfigure2}. As shown in Figs.~\ref{Sfigure2}(a), \ref{Sfigure2}(e), and \ref{Sfigure2}(i), when pumping from port 1, $I_{\rm A}$ exhibits significant irregularities in the time domain, and its motion trajectories are intertwined, forming a series of solid ribbon structures on the phase plane, and its frequency spectrum appears continuous. All of these phenomena indicate that the device is operating in chaotic mode, which means that even if the equations of motion are deterministic, we still cannot accurately predict the future state of the device.

However, if we switch the pumping port from 1 to 2 and keep other parameters constant, then the operating mode of the device will change dramatically. As shown in Figs.~\ref{Sfigure2}(b) and \ref{Sfigure2}(j), $I_{\rm A}$ is in a stable periodic oscillation state, which can be seen more intuitively from the frequency spectrum (discrete peaks occur only at integer multiples of $\Omega$). Although the motion trajectories in phase space appear somewhat disorganized~[see Fig.~\ref{Sfigure2}(f)], the obvious limit-cycle characteristics can still be discerned, which also proves that the device is operating in an ordered mode. In summary, the Lorentz reciprocity of our device has been broken (except for some special cases, which will be discussed in detail later), and the same conclusion can be drawn from Figs.~\ref{Sfigure2}(c), \ref{Sfigure2}(d), \ref{Sfigure2}(g), \ref{Sfigure2}(h), \ref{Sfigure2}(k), and \ref{Sfigure2}(l).

\begin{figure}[t]
	\centering
	\includegraphics[width= 12 cm]{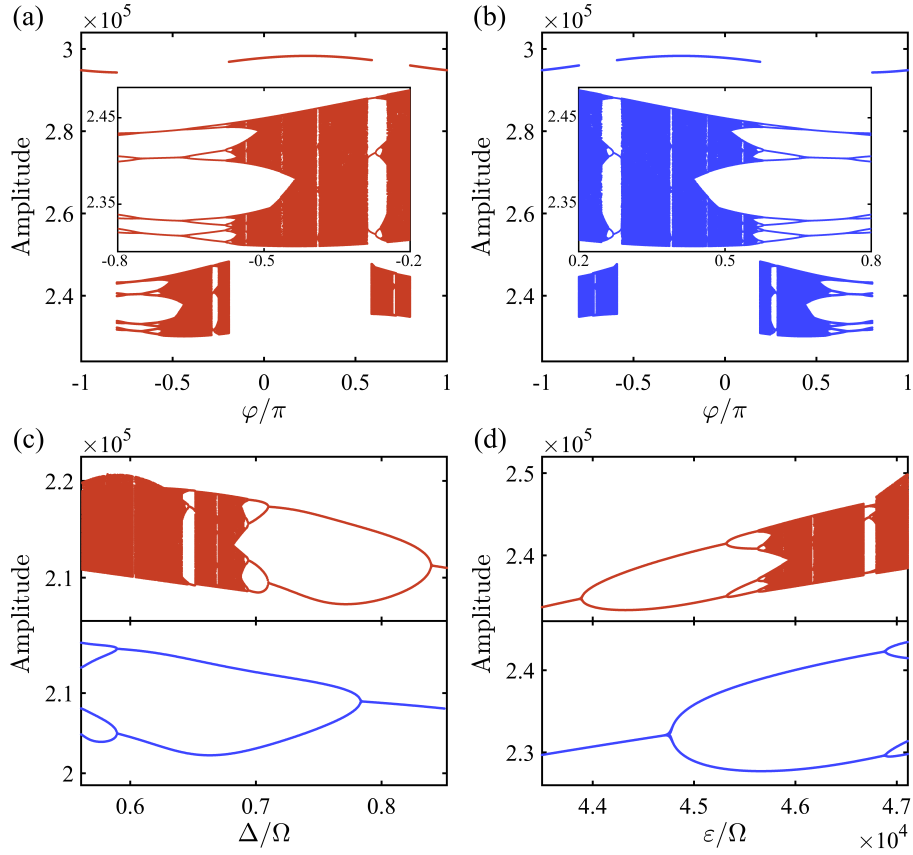}
	\caption{Bifurcation diagrams of the mechanical oscillation amplitude for pumping from port 1 (red) and port 2 (blue), under different parameter conditions. Here, for panels (a) and (b), $|\xi|/\Omega = 3$ and both insets are enlarged views; for panel (c), $\Delta=\omega_{\rm A}-\omega=\omega_{\rm B}-\omega$, $\varepsilon/\Omega = 5.4\times10^{\;\;\!\!\!4}$, $|\xi|/\Omega = 5$, and $\varphi/\pi=0.5$; for panel (d), $\varphi/\pi=0.5$ and $|\xi|/\Omega = 10$; other parameters in all panels are the same as in Fig.~\ref{Sfigure2}.}
\label{Sfigure3}
\vspace{-0.3cm}
\end{figure}

Another important finding is that if the hopping phase is transformed from $\varphi$ to $-\;\!\varphi$ while switching the pumping port, then the operating state of the device remains unchanged (compare the left and right parts of Fig.~\ref{Sfigure2}). To fully demonstrate this, in Figs.~\ref{Sfigure3}(a) and \ref{Sfigure3}(b), we plot bifurcation diagrams (vs $\varphi$) under two pumping conditions. Now, we just need to count the number of bifurcations to distinguish different operating modes~\cite{PRL114013601}. More specifically, one, multiple, and infinite scatter points on the diagram indicate the self-induced oscillation, period-doubling, and chaotic modes of the device, respectively. A remarkable phenomenon is that no matter which port is pumped, the device can be switched between regular and chaotic operating modes by adjusting $\varphi$. However, such switching processes are not completely independent for different pumping ports. Comparing the results in Figs.~\ref{Sfigure3}(a) and \ref{Sfigure3}(b), it can be easily found that they satisfy chiral symmetry on the $\varphi\;\!\;\!\!$-$\rm axis$, which is particularly evident on the two insets. Therefore, we conclude that the proposed device supports a chiral chaotic operating mode.

In addition, it is also worth noting that in certain parameter regions, chaos can only be generated along one specific pumping direction. To show this nonreciprocal chaotic operating mode more clearly, we plot the representative results in Figs.~\ref{Sfigure3}(c) and \ref{Sfigure3}(d). Both figures demonstrate that the response of our device to external pumping (frequency $\omega$ and amplitude $\varepsilon$) depends on the operating port. As shown in the red parts of Figs.~\ref{Sfigure3}(c) and \ref{Sfigure3}(d), when pumping from port 1, adjusting $\omega$ or $\varepsilon$ can control the device to operate in a desired mode, either chaotic or regular. However, after switching the pumping port, as shown in the blue parts of Figs.~\ref{Sfigure3}(c) and \ref{Sfigure3}(d), device will be restricted to the regular operating regime, which means that all chaotic modes are no longer available. Similar nonreciprocal behavior exists for other parameter regions as well. Thus, the designed device can be applied to realize multiple types of nonreciprocal signal modulation, which indicates its potential applications in communication networks and quantum information processing~\cite{Nature4531023,Nature541473}.

\subsection{Analysis for gauge transformations and time-reversal symmetry}

Physically, the chirality in our system originates from the manipulating of degeneracy or time-reversal symmetry of CW and CCW modes by tips. We now demonstrate this from the perspective of gauge transformations. According to the standard analysis method, if our system maintains time-reversal symmetry, then its Hamiltonian needs to satisfy the transformation relation $\hat{\Theta} \hat{H} \hat{\Theta}^{-1}=\hat{H}$, where $\hat{\Theta}$ represents the antiunitary and antilinear time-reversal operation. Furthermore, the transformation relations of optical and mechanical operators under time-reversal reads
\begin{gather}
	\hat{\Theta} \hat{a}_{\rm CW} \hat{\Theta}^{-1}=e^{i\theta_{1}}\hat{a}_{\rm CW},\quad
	\hat{\Theta} \hat{a}_{\rm CCW} \hat{\Theta}^{-1}=e^{i\theta_{2}}\hat{a}_{\rm CCW},\quad\hat{\Theta}\hat{q}\;\!\;\!\!\hat{\Theta}^{-1}=\hat{q},\label{S22}
\end{gather}
\begin{gather}
	\hat{\Theta} \hat{b}_{\rm CW} \hat{\Theta}^{-1}=e^{i\theta_{3}}\hat{b}_{\rm CW},\quad \hat{\Theta} \hat{b}_{\rm CCW} \hat{\Theta}^{-1}=e^{i\theta_{4}}\hat{b}_{\rm CCW},\quad \hat{\Theta} \hat{p}\;\!\;\!\! \hat{\Theta}^{-1}=-\;\;\!\!\!\hat{p}+\nabla\vartheta(\hat{q}),\label{S23}
\end{gather}
where $\{\theta_{1},\theta_{2},\theta_{3},\theta_{4},\vartheta(\hat{q})\}$ represent the linear gauge phases. One can easily verify that by virtue of the gauge phases, the following relations always hold
\begin{align}
\hat{\Theta}(\hat{a}^{\dag}_{\rm CW}\hat{a}_{\rm CW}+\hat{a}^{\dag}_{\rm CCW}\hat{a}_{\rm CCW})\hat{q}\;\;\!\!\! \hat{\Theta}^{-1}&=(\hat{a}^{\dag}_{\rm CW}\hat{a}_{\rm CW}+\hat{a}^{\dag}_{\rm CCW}\hat{a}_{\rm CCW})\hat{q},\\[3pt]
\hat{\Theta}(\hat{a}^{\dag}_{\rm CW}\hat{a}_{\rm CW}+\hat{a}^{\dag}_{\rm CCW}\hat{a}_{\rm CCW})\hat{\Theta}^{-1}&=\hat{a}^{\dag}_{\rm CW}\hat{a}_{\rm CW}+\hat{a}^{\dag}_{\rm CCW}\hat{a}_{\rm CCW},\\[3pt]
\hat{\Theta}(\hat{b}^{\dag}_{\rm CW}\hat{b}_{\rm CW}+\hat{b}^{\dag}_{\rm CCW}\hat{b}_{\rm CCW})\hat{\Theta}^{-1}&=\hat{b}^{\dag}_{\rm CW}\hat{b}_{\rm CW}+\hat{b}^{\dag}_{\rm CCW}\hat{b}_{\rm CCW},\\[3pt]
\hat{\Theta}(\hat{p}^2+\hat{q}^{2}) \hat{\Theta}^{-1}=\hat{p}^{2}\,\;\;\!\!\!+&\,\;\;\!\!\!\hat{q}^2, \quad {\rm for}\; \nabla\vartheta(\hat{q})=0.
\end{align}
However, this is not the case for interaction terms between optical modes. According to Eqs.~(\ref{S22}) and (\ref{S23}), we get the following transformation relations
\begin{align}
	\hat{\Theta}(\hat{a}^{\dag}_{\rm CW}\hat{b}_{\rm CCW}+\hat{a}_{\rm CW}\hat{b}_{\rm CCW}^{\dag})\hat{\Theta}^{-1}&=e^{i(\theta_{4}-\theta_{1})}\hat{a}^{\dag}_{\rm CW}\hat{b}_{\rm CCW}+e^{i(\theta_{1}-\theta_{4})}\hat{a}_{\rm CW}\hat{b}_{\rm CCW}^{\dag},\label{S28}\\[3pt]
	\hat{\Theta}(\hat{a}^{\dag}_{\rm CCW}\hat{b}_{\rm CW}+\hat{a}_{\rm CCW}\hat{b}_{\rm CW}^{\dag})\hat{\Theta}^{-1}&=e^{i(\theta_{3}-\theta_{2})}\hat{a}^{\dag}_{\rm CCW}\hat{b}_{\rm CW}+e^{i(\theta_{2}-\theta_{3})}\hat{a}_{\rm CCW}\hat{b}_{\rm CW}^{\dag},\label{S29}\\[3pt]
	\hat{\Theta}(\hat{a}^{\dag}_{\rm CW}\hat{a}_{\rm CCW}+\hat{a}_{\rm CW}\hat{a}_{\rm CCW}^{\dag})\hat{\Theta}^{-1}&=e^{i(\theta_{2}-\theta_{1})}\hat{a}^{\dag}_{\rm CW}\hat{a}_{\rm CCW}+e^{i(\theta_{1}-\theta_{2})}\hat{a}_{\rm CW}\hat{a}_{\rm CCW}^{\dag},\label{S30}\\[3pt]
	\hat{\Theta}(e^{i\varphi}\hat{b}^{\dag}_{\rm CW}\hat{b}_{\rm CCW}+e^{-i\varphi}\hat{b}_{\rm CW}\hat{b}_{\rm CCW}^{\dag}) \hat{\Theta}^{-1}&=e^{i(\theta_{4}-\theta_{3}-\varphi)}\hat{b}^{\dag}_{\rm CW}\hat{b}_{\rm CCW}+e^{i(\theta_{3}-\theta_{4}+\varphi)}\hat{b}_{\rm CW}\hat{b}_{\rm CCW}^{\dag}.\label{S31}
\end{align}
Therefore, the gauge phases that make the Hamiltonian satisfy time-reversal transformation relation must meet
\begin{align}
	\theta_{4}-\theta_{1}=2\;\;\!\!\!\pi\;\;\!\!\!\mathbb{Z}_{1},\quad\theta_{3}-\theta_{2}=2\;\;\!\!\!\pi\;\;\!\!\!\mathbb{Z}_{2},\quad\theta_{2}-\theta_{1}=2\;\;\!\!\!\pi\;\;\!\!\!\mathbb{Z}_{3},\quad\theta_{3}-\theta_{4}+2\;\;\!\!\!\varphi=2\;\;\!\!\!\pi\;\;\!\!\!\mathbb{Z}_{4},\label{S32}
\end{align}
where $\{\mathbb{Z}_{1}, \mathbb{Z}_{2}, \mathbb{Z}_{3}, \mathbb{Z}_{4}\}$ represent arbitrary integers. Further processing Eq.~(\ref{S32}), we obtain the necessary and sufficient condition for the system to hold time-reversal symmetry, that is, $\varphi=\pi\;\;\!\!\!\mathbb{Z}$ ($\mathbb{Z}$ denotes arbitrary integer). This implies that the extra hopping phase $\varphi$ acquired by photons during tips-induced backscattering processes can also be applied as a means to regulate the dynamics. For nontrivial $\varphi$, i.e., $\varphi\neq\pi\;\;\!\!\!\mathbb{Z}$, as shown in Eqs.~(\ref{S28})--(\ref{S31}), its value is opposite for the two reverse hopping routes, which is the root cause of chirality.

\subsection{Steady-state light intensity distributions}

Although the analytical expression for chiral chaos is unavailable, we can analytically obtain the steady-state light intensity distributions, which also reveal the origin of chirality. For steady-state, Eqs.~(\ref{S2})--(\ref{S6}) can be rewritten as
\begin{align}
	0=&-i(\Delta_{\rm A}-Gq_{s}-i\kappa)a_{\rm CW}^{s}+i\eta a_{\rm CCW}^{s}+iJb_{\rm CCW}^{s}+\varepsilon_{1},\\[3pt]
	0=&-i(\Delta_{\rm A}-Gq_{s}-i\kappa)a_{\rm CCW}^{s}+i\eta a_{\rm CW}^{s}+iJb_{\rm CW}^{s}+\varepsilon_{2},\\[3pt]
	0=&-i(\Delta_{\rm B}-i\gamma)b_{\rm CW}^{s}+i|\xi|e^{i\varphi}b_{\rm CCW}^{s}+iJa_{\rm CCW}^{s},\\[3pt]
	0=&-i(\Delta_{\rm B}-i\gamma)b_{\rm CCW}^{s}+i|\xi|e^{-i\varphi}b_{\rm CW}^{s}+iJa_{\rm CW}^{s},\\[3pt]
	0=&\ \Omega\:\!\:\! p_{s},\quad 0=-\ \!\;\;\!\!\!\Omega\:\! q_{s}+G(|a_{\rm CW}^{s}|^{2}+|a_{\rm CCW}^{s}|^{2})-\Gamma\:\! p_{s}.
\end{align}
Solving the above equations, we get the following results
\begin{align}
	a_{\rm CW}^{s}&=-i\dfrac{\varepsilon_{2}(\eta+Fe^{-i\varphi})+\varepsilon_{1}\widetilde{\Delta}_{\rm A}}{\widetilde{\Delta}_{\rm A}^{2}-(\eta+Fe^{-i\varphi})(\eta+Fe^{i\varphi})},\quad a_{\rm CCW}^{s}=-i\dfrac{\varepsilon_{1}(\eta+Fe^{i\varphi})+\varepsilon_{2}\widetilde{\Delta}_{\rm A}}{\widetilde{\Delta}_{\rm A}^{2}-(\eta+Fe^{-i\varphi})(\eta+Fe^{i\varphi})},\label{S38}\\[3pt]
	b_{\rm CW}^{s}&=\frac{J|\xi|e^{i\varphi}a_{\rm CW}^{s}+J(\Delta_{\rm B}-i\gamma)a_{\rm CCW}^{s}}{(\Delta_{\rm B}-i\gamma)^2-|\xi|^2},\quad b_{\rm CCW}^{s}=\frac{J|\xi|e^{-i\varphi}a_{\rm CCW}^{s}+J(\Delta_{\rm B}-i\gamma)a_{\rm CW}^{s}}{(\Delta_{\rm B}-i\gamma)^2-|\xi|^2},\label{S39}\\[3pt]
	&\qquad \qquad \qquad \qquad \quad p_{s}=\;\!0,\quad q_{s}=\dfrac{G}{\Omega}(|a_{\rm CW}^{s}|^{2}+|a_{\rm CCW}^{s}|^{2}),
\end{align}
in which
\begin{gather}
	\widetilde{\Delta}_{\rm A}=\Delta_{\rm A}-\dfrac{G^2}{\Omega}(|a_{\rm CW}^{s}|^{2}+|a_{\rm CCW}^{s}|^{2})-i\kappa-\dfrac{F(\Delta_{\rm B}-i\gamma)}{|\xi|},\quad F=\dfrac{J^2|\xi|}{(\Delta_{\rm B}-i\gamma)^2-|\xi|^2}.\tag{S41}
\end{gather}
Therefore, the steady-state light intensities in resonator A $(I_{\rm A}^{s}=|a_{\rm CW}^{s}|^2+|a_{\rm CCW}^{s}|^2)$ corresponding to pumping from port 1 and port 2 can be expressed as
\begin{gather}
 I_{\rm A,1}^{s}=g(|\widetilde{\Delta}_{\rm A}|^2+|\eta+Fe^{i\varphi}|^2),\quad {\rm \ \;\;\!\!\!\;\;\!\!\!\;\;\!\!\!\;\;\!\!\!\;\;\!\!\!pumping\ from\ port\ 1},\tag{S42}\\[3pt]
 I_{\rm A,2}^{s}=g(|\widetilde{\Delta}_{\rm A}|^2+|\eta+Fe^{-i\varphi}|^2),\quad {\rm pumping\ from\ port\ 2}.\tag{S43}
\end{gather}
Here, $g=\varepsilon^2/|\widetilde{\Delta}_{\rm A}^{2}-(\eta+Fe^{-i\varphi})(\eta+Fe^{i\varphi})|^2$ and we have omitted $G^2(|a_{\rm CW}^{s}|^2+|a_{\rm CCW}^{s}|^2)/\Omega$ (since $G\ll\Omega)$. Similarly, the steady-state light intensities in resonator B $(I_{\rm B}^{s}=|b_{\rm CW}^{s}|^2+|b_{\rm CCW}^{s}|^2)$ can also be obtained, but the specific forms are too cumbersome to list here. For the convenience of comparison, we further define $\delta I_{\rm A}^{s}=I_{\rm A,1}^{s}-I_{\rm A,2}^{s}$ as the steady- state light intensity difference in resonator A caused by pumping from different ports. Obviously, when time-reversal symmetry holds, i.e., $\varphi=\pi\;\;\!\!\!\mathbb{Z}$, $\delta I_{\rm A}^{s}=0$, and when time-reversal symmetry breaks, $\delta I_{\rm A}^{s}=g\;\;\!\!\!(|\eta+Fe^{i\varphi}|^2-|\eta+Fe^{-i\varphi}|^2)$. One can easily verify that $I_{\rm A,1}^{s}\Rightarrow$ $I_{\rm A,2}^{s}$ and $\delta I_{\rm A}^{s}\Rightarrow-\;\;\!\!\!\delta I_{\rm A}^{s}$ under the transformation $\varphi\Rightarrow-\;\;\!\!\!\varphi$. Such a transformation relationship conforms the chiral symmetry of system dynamics on the $\varphi\;\;\!\!\!$-axis. Finally, it is also worth mentioning that the steady-state solutions predict several other useful operating modes. For example, scattering induced by intrinsic defects can be greatly suppressed by properly selecting parameters, which is essential for constructing robust channels. As shown in Eq.~(\ref{S38}), if $\eta+Fe^{i\varphi}\rightarrow0$, then the CCW mode in resonator A will be completely suppressed when pumping from port 1. The same phenomenon also exists for the CW mode. Detailed discussion on this aspect is beyond the scope of this Letter and will be undertaken in the future.

\subsection{Explanations of the structural parameters defined in the main text}

\begin{figure}[t]
	\centering
	\includegraphics[width= 16 cm]{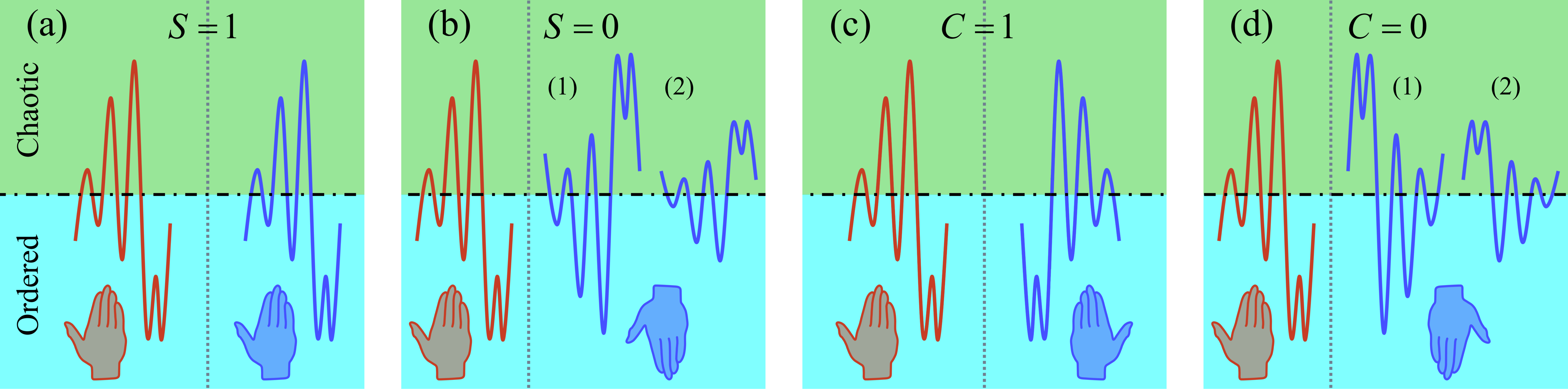}
	\caption{Geometric interpretations for the structural parameters $S$ and $C$. Here, the black dash-dotted lines represent $\lambda_{\rm max}=0$, and the green and cyan regions separated by it represent $\lambda_{\rm max}>0$ (chaotic) and $\lambda_{\rm max}<0$ (ordered), respectively. In all panels, the red and blue curves represent the maximal Lyapunov exponent spectra for the corresponding extreme case (S, C reach 0 or 1), where in panels (b) and (d), two different types of blue curves are presented as (1) and (2), since they both meet the requirements for complete asymmetry or achirality.  
 The different positions of hands at the bottoms are indicators for the structural features of the corresponding S or C.}
	\label{Sfigure4}
	\vspace{-0.3cm}
\end{figure}

We now provide detailed explanations and descriptions for the structural parameters $S$ and $C$ defined in the main text. First, we briefly review the Lyapunov exponents. (i) Lyapunov exponents describe the degree of separation over time between neighboring trajectories in phase space. (ii) They have been widely accepted as metrics to characterize the dynamics of nonlinear systems, and many reliable calculation algorithms have been proposed~\cite{Meccanica1980}. (iii) In general, their values are extremely sensitive to changes in parameters of the differential or map equations. (iv) The sign of the maximal Lyapunov exponent $\lambda_{\rm max}$ is opposite for ordered $(\lambda_{\rm max}<0)$ and chaotic $(\lambda_{\rm max}>0)$ regimes, which ensures its excellent distinguishing ability~\cite{PRL114013601}. Second, we clarify the underlying assumptions implicit in the structural parameters. (I) If the evolution of system dynamics (order-chaos transitions, bifurcations, catastrophes, etc.) exhibits exactly the same properties under two different sets of conditions, then the two sets of conditions are considered to be equivalent and the corresponding evolution processes are ideally symmetrical. (II) If the evolution of system dynamics under two different sets of conditions are mirror images of each other, then the two sets of conditions are considered as enantiomers and the corresponding evolution processes are ideally chiral~\cite{AM2022}.

Combining the Lyapunov exponents and the above-mentioned basic assumptions, we propose the following metrics to measure the symmetry and chirality of any two routes to chaos. Define $x$ as the condition under which the system is in state $U$ with a maximal Lyapunov exponent $\lambda_{\rm max}$. Assign $N$ such conditions into set $\{x_{i}\}$ according to certain rules, we can obtain an evolution process of system dynamics $\{U_{i}\}$ and the corresponding maximal Lyapunov exponent array $\{\lambda_{\rm max}^{i}\},$ $i=1,2,\cdots\;\;\!\!\!\!,N$. Similarly, we create another $N$-element condition set $\{y_{i}\}$, and obtain another evolution process $\{V_{i}\}$ and a new maximal Lyapunov exponent array $\{\Lambda^{i}_{\rm max}\},$ $i=1,2,\cdots\;\;\!\!\!\!,N$. Then, we define the following structural parameters
\begin{gather}	
	S=1-\dfrac{1}{N}\sum_{i=1}^{N}\dfrac{|\lambda^{i}_{\rm max}-\Lambda^{i}_{\rm max}|}{|\lambda^{i}_{\rm max}|+|\Lambda^{i}_{\rm max}|},\quad C=1-\dfrac{1}{N}\sum_{i=1}^{N}\dfrac{|\lambda^{i}_{\rm max}-\Lambda^{N-i+1}_{\rm max}|}{|\lambda^{i}_{\rm max}|+|\Lambda^{N-i+1}_{\rm max}|}.
	\tag{S44}
	\label{S44}
\end{gather}

Obviously, both parameters take values between $0$ and $1$ (inclusive). According to the previous assumptions, if two evolution processes are ideally symmetrical, i.e., $U_{i}=V_{i}$ and $\lambda^{i}_{\rm max}=\Lambda^{i}_{\rm max}$, then $S=1$~[see Fig.~\ref{Sfigure4}(a)]. On the contrary, if two processes are completely asymmetrical, that is, the operating states represented by each pair of $U_{i}$ and $V_{i}$ are completely different (one is ordered and the other is chaotic), then the signs of $\lambda^{i}_{\rm max}$ and $\Lambda^{i}_{\rm max}$ must be opposite, resulting in $S=0$~[see Fig.~\ref{Sfigure4}(b)]. Note that the conditions for $S=1$ are stricter than those for $S=0$. As shown in Figs.~\ref{Sfigure4}(a) and \ref{Sfigure4}(b), for the process described by the red curve, there is only one ideal symmetrical process, while there are countless completely asymmetrical processes, e.g., (1) and (2) in Fig.~\ref{Sfigure4}(b). For the general case, $S\in(0,1)$, which can be understood as the average degree of symmetry between two evolution processes. In the same way, if two processes are ideally chiral, i.e., $U_{i}=V_{N-i+1}$ and $\lambda^{i}_{\rm max}=\Lambda^{N-i+1}_{\rm max}$, then $C=1$~[see Fig.~\ref{Sfigure4}(c)]. For completely achiral, one of $U_{i}$ and $V_{N-i+1}$ is ordered and the other is chaotic, which means the signs of $\lambda^{i}_{\rm max}$ and $\Lambda^{N-i+1}_{\rm max}$ are opposite, resulting in $C=0$~[see Fig.~\ref{Sfigure4}(d)]. The conditions to be satisfied for $C = 1$ are also stricter than those for $C = 0$. As shown in Figs.~\ref{Sfigure4}(c) and \ref{Sfigure4}(d), for the process described by the red curve, there is only one ideal chiral process, while there are countless completely achiral processes, e.g., (1) and (2) in Fig.~\ref{Sfigure4}(d). For other cases, $C\in(0,1)$, which can be understood as the average degree of chirality between two evolution processes. In conclusion, the proposed structural parameters can properly quantify the symmetry and chirality between routes to different chaos~\cite{RevModPhys71}.

\section{Details of chiral chaos for sensitive sensing}

The core principle of the sensing method we proposed, i.e., constructing the sensing window, has been explained in the main text. Below, we take the detection of monochromatic optical (weak) signal as an example to test the actual performance of sensing windows. More complicated cases, such as signals with a certain bandwidth or signals with background noise, will be discussed in our follow-up work. The specific form of the weak signal is
\begin{equation}
	s(\tau)=\;\;\!\!\!\delta\varepsilon\;\;\!\!\!{\rm cos}\;\;\!\!\![(\omega+\delta\omega)\tau+\theta\;\;\!\!\!],\quad \delta\varepsilon=\sqrt{\dfrac{\kappa_{0}\;\;\!\!\!p_{s}}{\hbar(\omega+\delta\omega)}}\;\;\!\!\!,
	\tag{S45}\label{S45}
\end{equation}
where $\delta\varepsilon$, $\omega+\delta\omega$, and $p_{s}$ represent the amplitude, frequency, and power of the weak signal, respectively. Here, $\omega$ and $\varepsilon$ $(\varepsilon\gg\delta\varepsilon)$ denote the frequency and amplitude of the pumping field, $\theta$ and $\delta\omega$ represent the initial phase difference and detuning between the signal and the pumping field, respectively, $\kappa_{0}$ is the taper-resonator coupling rate, and $\tau$ ($\tau=\Omega\;\;\!\!\!t$) is the extended time scale. To simplify subsequent discussions and facilitate experimental verification, we assume that the weak signal is directly loaded into the pumping field. Then, the equivalent Hamiltonian is described as
\begin{equation}
		\hat{H}=\left\{
		\begin{split}
			&i\hbar\!\!\:\left[\varepsilon+\delta\varepsilon\;\;\!\!\! e^{-i(\delta\omega \tau\;\!+\;\!\theta)}\;\!\!\right]\!\!\:\hat{a}_{\rm CW}^{\dag}+{\rm H.c.},\ {\rm input\ from\ port\ 1},
			\\[3pt]
			&i\hbar\!\!\:\left[\varepsilon+\delta\varepsilon\;\;\!\!\! e^{-i(\delta\omega \tau\;\!+\;\!\theta)}\;\!\!\right]\!\!\:\hat{a}_{\rm CCW}^{\dag}+{\rm H.c.},\ {\rm input\ from\ port\ 2},	
		\end{split}
		\right.
	\tag{S46}\label{S46}
\end{equation}
where a rotating frame at frequency $\omega$ and the rotating-wave approximation have been applied.

For a more intuitive comparison, we stimulate a conventional chaos-based sensor~\cite{IEEE464401999} (based on order-chaos phase transition) using the sensing results when inputting from port 2 of our device. Actually, switching the input port has no effect on the sensing performance of conventional reciprocal chaos-based sensors. However, the situation is completely different in our chiral chaotic sensor. After constructing a sensing window, we need to superimpose the contributions from the two input ports to achieve the final sensing result.

\begin{figure}
	\centering
	\includegraphics[width=12cm]{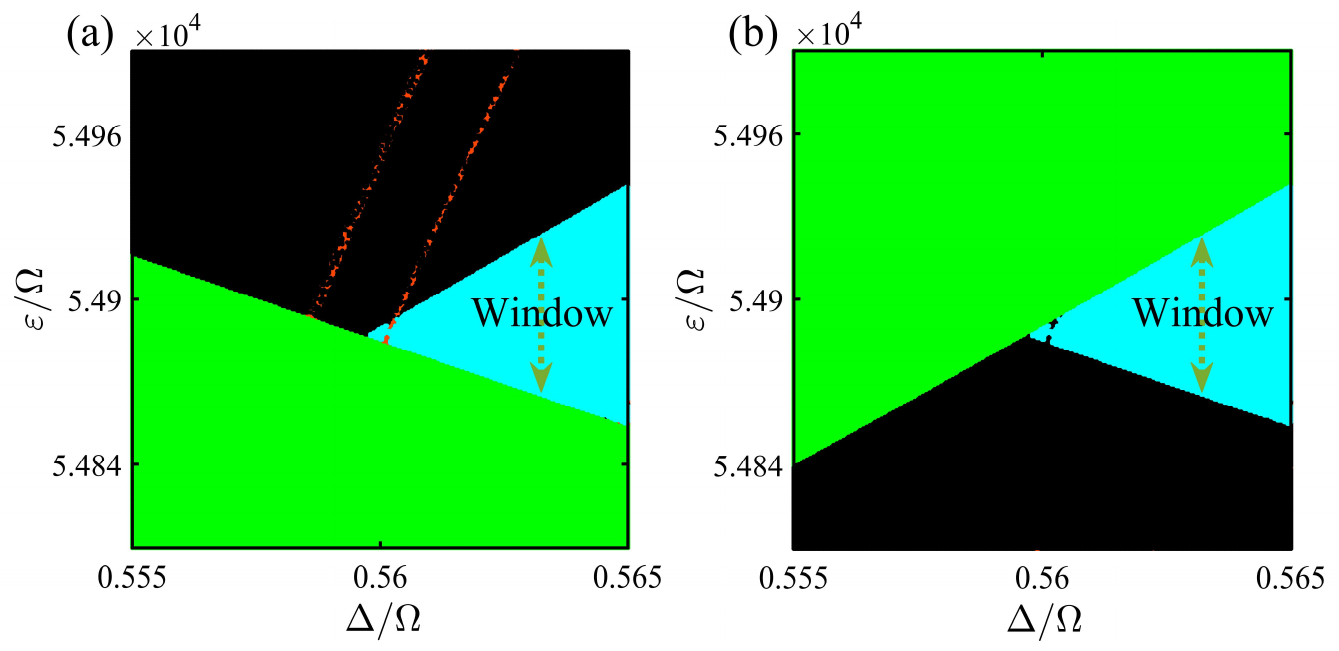}
	\caption{Panels (a) and (b) are phase diagrams of the system within the selected parameter regions when pumping from port 1 and port 2, respectively. Here, the green, red, and black regions represent the system locates in the self-induced oscillations, period-doubling bifurcations, and chaotic phases, respectively. The cyan region represents a special type of chaotic phase, where the system is always in the chaotic phase no matter which port is pumped. The parameters are $\Delta_{\rm A}=\Delta_{\rm B}=\Delta$, $\delta\varepsilon=0$, $\varphi/\pi = $ $0.5$, $|\xi|/\Omega= 5$, and other parameters are the same as in Fig.~\ref{Sfigure2}.}
	\label{Sfigure5}
	\vspace{-0.3cm}
\end{figure}

Before testing, we need to construct the required sensing window. As shown in the phase diagrams in Fig.~\ref{Sfigure5}, the operating states of the device are different when pumping from different ports. It should be noted that a special type of region exists on the phase diagrams, that is, the sensing window region. In such a region, the system dynamics is sensitive to detected signals, and this sensitivity is not affected by the randomness in certain parameters of the signals. For example, the cyan region on the phase diagrams is a sensing window region. In this region, the system is always in the chaotic state no matter which port is pumped. Furthermore, for a resonant ($\delta\omega=0$) or near-resonant ($\delta\omega/\omega\ll1$) signal with amplitude $|\delta\varepsilon|$, even if the value of $\theta$ is random, the system still can undergo a chaos-to-order phase transition upon pumping from one of the ports. For $\theta\in(0,\pi/2)$, the chaos-to-order phase transition can be induced when input from port 2, for $\theta\in(\pi/2,\pi)$, the chaos-to-order phase transition can be induced when input from port 1. As illustrated by the bidirectional arrows in Fig.~\ref{Sfigure5}, each vertical parameter bar in the cyan region can be designed as a sensing window, and the half-width and center position of the window are determined by the upper and lower boundaries. It is obvious that the characteristics of sensing windows can be adjusted smoothly over a wide range. Therefore, the application scope of such sensing window region can be greatly broadened. Narrow windows are suitable for sensitive sensing, while wide windows are suitable for threshold alarms~\cite{J.S.Wilson2005}. In our main text, the sensing window region constructed on the $(\varepsilon,\omega)$ plane is located in the cyan region in Fig.~\ref{Sfigure5}, and the operating point $F$ is at $\varepsilon/\Omega=54887$ and $\Delta/\Omega=0.5598$.

\end{document}